\documentclass[12pt]{article}
\usepackage{amsmath}
\usepackage{graphicx,psfrag,epsf}
\usepackage{enumerate}
\usepackage{natbib}
\usepackage{xcolor}
\usepackage{amssymb}
\usepackage{graphicx}
\usepackage{multicol}
\usepackage{multirow}
\usepackage{float}
\usepackage{listings}
\usepackage{comment}
\usepackage{bbm}
\usepackage{authblk} 

\usepackage{url} 
\newcommand{\am}[1]{{\textcolor{black}{{{}}#1}}{}}
\newcommand{\davidsal}[1]{{\textcolor{black}{{{}}#1}}{}}
\newcommand{\amrev}[1]{{\textcolor{black}{{{}}#1}}{}}

\usepackage[left=2.5cm, right=2.5cm, top=2cm, bottom=2cm]{geometry}
\usepackage[]{algorithm2e}
\usepackage{changepage}

\begin{document}

\title{\bf Bayesian modelling and computation utilising directed cycles in multiple network data}

\author[1]{Anastasia Mantziou}
\author[2]{Sally Keith}
\author[2]{David M.P. Jacoby}
\author[3]{Sim\'{o}n Lunag\'{o}mez}
\author[4]{Robin Mitra}

\affil[1]{Department of Statistics, University of Warwick, Coventry, CV4 7AL, U.K.}
\affil[2]{Lancaster Environment Centre, Lancaster University, Lancaster, La1 4YQ, U.K.}
\affil[3]{Department of Statistics, ITAM, Rio Hondo, M\'{e}xico, 01080}
\affil[4]{Department of Statistical Science, University College London, Gower Street, London, WC1E 6BT, U.K.}

\date{}

\maketitle

\vspace{-0.7cm}

\begin{abstract}
Modelling multiple network data is crucial for addressing a wide range of applied research questions. However, there are many challenges, both theoretical and computational, to address.  Network cycles are often of particular interest in many applications; \davidsal{for example in ecology a largely unexplored area has been how to incorporate network cycles within the inferential framework in an explicit way. } 
The recently developed Spherical Network Family of models (SNF) offers a flexible formulation for modelling multiple network data that permits any type of metric. This has opened up the possibility to formulate network models that focus on network properties hitherto not possible or practical to consider. In this article we propose a novel network distance metric that measures similarities between networks with respect to their cycles, and incorporates this within the SNF model to allow inferences that explicitly capture information on cycles. These network motifs are of particular interest \davidsal{in ecological studies aimed at understanding competitive and hierarchical interactions.} We further propose a novel computational framework to allow posterior inferences from the intractable SNF model for moderate-sized networks. Lastly, we apply the resulting methodology to a set of ecological network data studying aggressive interactions between species of fish. We show our model is able to make cogent inferences concerning the cycle behaviour amongst the species, and beyond those possible from a model that does not consider this network motif.
\end{abstract}

\noindent%
\vfill

\maketitle

\section{Introduction}
\label{s:intro}
In many fields, modelling network data is essential to answering the applied research questions of interest. In ecology, \davidsal{the interactive behaviour of different individuals or species} within a geographical area can be represented by a \textcolor{black}{directed} network with each species corresponding to a node and the edges representing interactions between species \citep{delmas2019analysing,mittelbachspecies}.
\davidsal{If these interactions are directed, such as aggression behaviour between individuals of different species, these networks can capture competitive species interactions at the ecosystem scale,} 
e.g. certain species vying over a particular food source or type of habitat. 
Each aggressive interaction can be represented through a directed edge (i.e. arrow) with the direction indicating which species is the aggressor and which is the \davidsal{recipient} respectively. 

Within ecology, a particular phenomenon of interest is where the aggressive interactions result in intransitive \davidsal{competition patterns, a set of cyclical interactions that results in no single dominant species, with different species winning out depending on the circumstances.} \am{Intransitive competition is of ecological importance as it is thought to promote species coexistence \citep{laird2006competitive}.} In a network these would be characterised through \textcolor{black}{directed} cycles \citep{koutrouli2020guide}. 

More generally, cycles reveal information about network topology \citep{maugis2017topology,fan2019towards}, and is a motif of interest in many applications beyond ecology. Examples include neuroscience, where the formation of cycles in a human brain network is crucial for human cognitive functions \citep{sizemore2018cliques}, and biology where RNAs forming covalently closed loop structures, called circular RNAs, have been associated with diseases such as cancer \citep{han2017circular}. 

Recent methodological developments permit data to be analysed where each observation is a network. In ecology, this arises when recording species' interactions across multiple different areas or sites. In this setting, the goal is to model the underlying mechanism that generates the multiple network data \davidsal{and to be able to directly compare networks across different spatial or temporal scales}. Recent studies have focused on the problem of modelling multiple network data utilising (a) a latent space framework \citep{gollini2016,durante2017,wang2019joint,nielsen2018multiple,arroyo2021inference}, (b) a measurement error process \citep{le2018estimating,newman2018estimating,peixoto2018reconstructing,mantziou2024bayesian,young2022clustering}, (c) distance functions \citep{lunagomez2020modeling,kolaczyk2020averages,ginestet2017hypothesis,josephs2023bayesian} \amrev{and (d) a Stochastic Block Model (SBM) structure \citep{josephs2023nested}.}
\amrev{Another study utilising an SBM structure for modelling multiplex networks is the study of \cite{amini2024hierarchical}. Multiplex networks \textcolor{black}{are} different to multiple networks as the former refers to networks observed at different layers with edges at each layer having a different interpretation of relation, while in multiple network data, the edges express the same type of relation across networks. In \cite{amini2024hierarchical}, the authors propose a hierarchical Stochastic Block Model (SBM) to recover communities of nodes at different network layers. }
However none of these models explicitly consider networks' cyclical properties in their \davidsal{formation, making it difficult to determine the underlying processes that sustain species interactions in the wild, or to measure how they might differ when drivers change.} 
Recent work utilising subgraph counts to test whether networks arise from a given distribution \citep{maugis2020testing} highlights how network properties, such as cycles, can be valuable in the analysis of network populations.

Distance-based models offer a way to encode networks' cycle information by incorporating this information in a metric measuring similarity between networks. 
There are a multitude of ways to define a similarity measure between different networks and for a review of these see \cite{donnat2018tracking}. However, none of these metrics explicitly consider cycles when measuring network dissimilarity.

In this article we propose a distance-based model for multiple network data that explicitly utilises the cycle information in the distance metric. Specifically, the metric we propose involves counting the number of uncommon cycles between the two networks, denoted as the symmetric difference, and combining this with the Hamming distance (both defined in the next section). The resulting metric is denoted as the Hamming-Symmetric difference (HS) distance metric. Enumerating cycles within a network is a computationally intensive task. \am{To deal with computational challenges in detecting large cycles, in this study we consider only directed cycles formed by three nodes, i.e. directed triangles. From an ecological perspective, three-node motifs are of interest \davidsal{as they indicate where on the transitive to intransitive continuum a species triad falls,
} 
with a directed triangle 
representing intransitive competition.} \textcolor{black}{There is thus an ecological, as well as mathematical, interest in studying directed triangles in network data.}

We adopt a Bayesian approach and utilise the Spherical Network Family (SNF) of models \citep{lunagomez2020modeling} to make posterior inferences as this gives us the flexibility to specify the distance metric of our choice, which in our setting is the HS distance. However, the computational challenges associated with fitting the SNF model are significant, with the model having an intractable normalising constant, which is a sum over the space of graphs. For a detailed review on methods for intractable distributions see \cite{park2018bayesian}. \textcolor{black}{Notably, for an directed network with $n$-nodes there are $2^{n(n-1)}$ possible networks, which means that even for a moderate-sized network with $n=20$ nodes there will be more than $2.46\times 10^{114}$ network configurations which are not practically possible to enumerate.}

The methodology proposed in \cite{lunagomez2020modeling} utilises \amrev{a diffusion distance metric and} the auxiliary variable implementation based on \cite{moller2006} to deal with the double intractability problem. \amrev{The choice of the diffusion distance metric in \cite{lunagomez2020modeling} is motivated by a neuroscience application, as diffusion distance can capture differences between networks with respect to how messages propagate through brain regions. However, for our ecological application, the specification of the diffusion distance metric would hinder ecological interpretation as it would not be clear how to translate message propagation in this setting. In our framework, we are interested in local changes in the structure of the networks with respect to edge flips and cycles differences rather than the global changes captured by the diffusion distance. The specification of our proposed HS distance metric allows for capturing such properties, however, the auxiliary variable technique formulated in \cite{lunagomez2020modeling}}
does not result in satisfactory performance for making posterior inferences in our setting. We thus develop an alternative computational framework to make posterior inferences through approximating the normalising constant using an Importance Sampler. This was inspired by an approach taken in \cite{vitelli2017}, albeit in a different setting with less computational challenges. 
The resulting modelling framework performs significantly better in making posterior inferences. 
\am{More details are given in Section \ref{sec53}}. We further evaluate our approach 
on
\davidsal{field data of competitive interactions between fish species at various reefs}
in the Indo-Pacific Ocean.

\textcolor{black}{While the seminal approach taken in \cite{lunagomez2020modeling} has opened up exciting possibilities for developing interpretable Bayesian models for multiple network data, and which our modelling framework is based on, we note a related earlier study 
\cite{banks1994metric}, which provides one of the first approaches in the literature for modelling multiple network data. 
\cite{banks1994metric} propose a model with the same functional form to the SNF model presented in \cite{lunagomez2020modeling}, and consider the Hamming distance metric to make inferences for multiple network data. The key differences with \cite{lunagomez2020modeling} are (i) the inferential framework in \cite{banks1994metric} is formulated for the Hamming distance and variants of it while \cite{lunagomez2020modeling} offer an inferential framework that allows the practitioner to specify a distance metric of their choice and (ii) \cite{banks1994metric} do not account for uncertainty quantification as in \cite{lunagomez2020modeling} who address this with a fully Bayesian framework.}

Our key contributions in this paper are thus three-fold. First, we propose a novel network distance metric, namely the HS distance, that has not been considered in the network literature. Second, we develop and implement a novel Markov Chain Monte Carlo (MCMC) scheme to make posterior inferences from the Spherical Network Family (SNF) model for multiple network data under the proposed HS distance metric. \textcolor{black}{Specifically, we introduce an Importance Sampling (IS) step to approximate the SNF model's intractable normalising constants within a Metropolis-Hastings (MH) algorithm. Third we utilise the modelling framework to infer cycle properties from a group of ecological networks studying aggressive interactions between species of fish.}

The remainder of this article is organised as follows. \textcolor{black}{Section \ref{sec:prelim} briefly reviews relevant fundamental network concepts, Section \ref{sec:application} describes our proposed metric as well as the ecological application that motivated its derivation. In Section \ref{sec:background} we give an overview of the SNF model and how \cite{lunagomez2020modeling} address the problem of the intractable normalising constant. In Section \ref{sec:meth} we present how we modify the computations to make posterior inferences for the SNF model and deal with the MCMC mixing issue, \amrev{along with simulation experiments for evaluating the performance of our method.} 
Section \ref{sec:realdata} applies our modelling framework to 
\davidsal{ecological data, specifically to quantify} aggressive interactions between \davidsal{coral-eating reef} species of fish.} Finally Section \ref{sec:conc} ends with some concluding remarks.

\section{Relevant network properties and preliminaries}\label{sec:prelim}

We represent a \am{directed} graph by $\mathcal{G}=(V,E)$, with $V= \{1,\ldots,n\}$ denoting the set of $n$ nodes and $E\subseteq \mathcal{E}_{n}$ denoting the set of edges in $\mathcal{G}$, with $\mathcal{E}_n = \{(i,j) \mid i,j\in V\}$. \am{Directed networks have ordered edges, such that $(i,j)$ is distinct to $(j,i)$.} \am{We note here that in this paper, we focus only on directed networks, however, our framework is easily adaptable to the simpler setting of undirected networks. }We use an $n\times n$ matrix, namely the adjacency matrix, to represent the presence and absence of edges in graph $\mathcal{G}$. Thus, the $(i,j)^{th}$ element of the adjacency matrix for a graph with binary edges is,
\begin{align}
A_{\mathcal{G}}(i,j)=
\begin{cases}
1,\text{ if an edge occurs \am{from} node i \am{to node} j,}\\
0,\text{ otherwise}.
\end{cases}
\end{align}
By $\mathcal{G}_{1},\ldots,\mathcal{G}_{N}$ we represent a population of $N$ \am{directed} graphs, with corresponding adjacency matrices $A_{\mathcal{G}_{1}},\ldots,A_{\mathcal{G}_{N}}$. We further assume that the networks in the population \am{have no} self-loops, and share the same set of $n$ \amrev{labelled} nodes \amrev{suggesting that the rows/columns of the adjacency matrices adhere to the same order}. We represent the space of graphs with $n$ nodes by $\{\mathcal{G}_{\mid n \mid}\}$, such that $\{\mathcal{G}_{\mid n \mid}\}=\{\mathcal{G}=(V,E):\mid V\mid=n\}$. Thus, the size of the space of \am{directed}, with no self-loops graphs is $\mid \{\mathcal{G}_{\mid n \mid}\}\mid=2^{n(n-1)}$.

A way to quantify similarities among networks is through the use of distance metrics which we denote by $d_{\mathcal{G}}(\cdot,\cdot)$. Two main types are: (a) structural distances that aim to capture similarities on edge-specific local properties of the graphs, and (b) to spectral distances that aim to capture similarities with respect to global properties of the graphs using a spectral representation \citep{donnat2018tracking}. A well-known structural distance metric is the \textit{Hamming distance}, that counts the not in common edges and non-edges between two graphs $\mathcal{G}_{k}$ and $\mathcal{G}_{l}$ for $k,l\in\{1,\ldots,N\}$. \am{The unnormalised Hamming distance between $\mathcal{G}_{k}$ and $\mathcal{G}_{l}$ is} defined as:
\begin{displaymath}
d_{H}(A_{\mathcal{G}_{k}},A_{\mathcal{G}_{l}})=\sum_{{i,j}} \mid A_{\mathcal{G}_{k}}(i,j)-A_{\mathcal{G}_{l}}(i,j)\mid.
\end{displaymath}

\begin{figure}[ht!]
\centering
\includegraphics[scale=0.15]{ 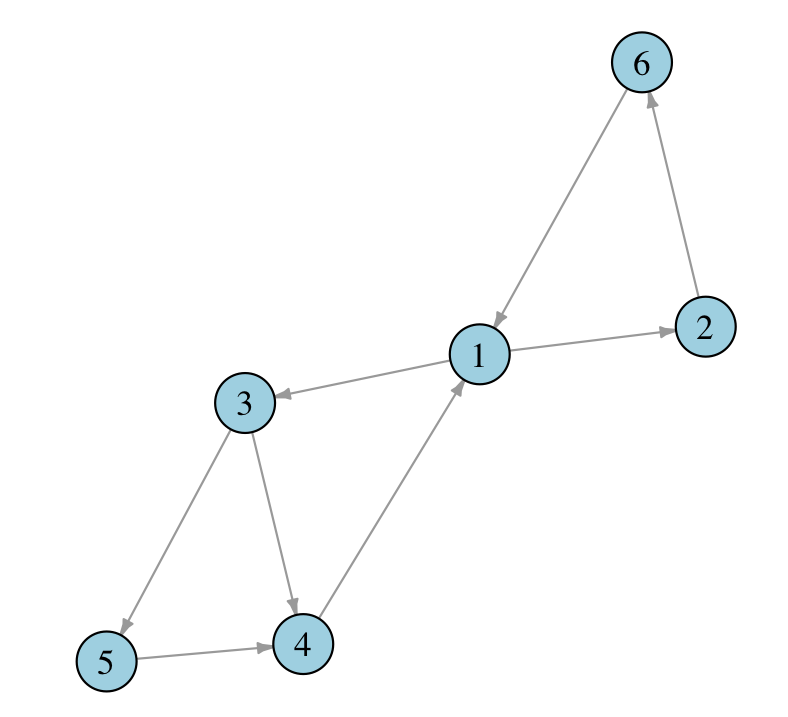}
\caption{Example of graph with \am{directed cycles} $\{1-2-6-1\}$, $\{1-3-5-4-1\}$ and $\{1-3-4-1\}$.}\label{toy}
\end{figure}

Networks are objects that can exhibit complex structures, thus the derivation of network properties such as the degree distribution is important for evaluating their characteristics. Network cycles are known to be crucial in revealing information about their topology \citep{maugis2017topology}. \am{ We acknowledge that terminology about motifs is not universal, and thus we use the following toy network to clarify what we mean by directed triangles which is the focus of our study. }\am{Firstly, a \textit{directed cycle} in a directed network} is a sequence of connected nodes in which the only repeated nodes are the first and the last node in the sequence. 
An illustrative example of an \am{directed} graph with \am{three directed cycles} is presented in Figure \ref{toy}. \am{We note here that $\{3,5,4\}$ does not form a directed cycle since there is an edge $(3,4)$ rather than $(4,3)$. The sequence of nodes $\{3,5,4\}$ and $\{1,3,4\}$ form triangles. In this study, we consider only directed cycles that form triangles (e.g. $\{1,3,4,1\}$), namely \textit{directed triangles}.}

\section{Ecological Application and proposed metric}
\label{sec:application}

Data have been collected on aggressive interactions between \davidsal{butterflyfish (genus Chaetodon)} on different coral reefs, \davidsal{across the Indo-Pacific region} \citep{keith2018synchronous}. We use a network representation, where nodes represent fish species and edges represent aggressive encounters. 

In Ecology, it is often of interest to identify competition structures among species that share resources\davidsal{, which enable particular ecological dynamics to be inferred e.g., hierarchical versus intransitive competition. Changes in competitive interactions can have broader effects, altering population dynamics, community structure, ecosystem function} \cite{envurl,mohd2019diversity,grether2017causes,kinlock2021uncovering}.
\textcolor{black}{In particular, the formation of directed cycles in graphs representing aggressive interactions among species point to intransitive competition patterns which are of particular interest ecologically \citep{sokhn2012identification,koutrouli2020guide}. 
Hence, a research question arising here is the following: How can we best use cyclical properties of networks, together with other network properties, to jointly analyse multiple network data, when the applied research questions pertain to \am{directed} cycles? We develop a Bayesian modelling and inferential framework to address this.}

An appealing way to answer this research question is with the SNF model \citep{lunagomez2020modeling} that infers a representative network in the population, determined through a user-specified distance metric. In addition, the SNF model involves a dispersion parameter that quantifies the level of dissimilarity between the network data and the network representative, with respect to the specified metric. The flexibility in the choice of the distance metric is a key motivating factor for developing a SNF model to analyse the data. 

As our interest lies in cycles formed in the network data, we propose a measure that captures information about dissimilarities between the cycles of two networks. Specifically, we propose a Hamming-Symmetric difference (HS) distance metric consisting of two parts:
\begin{enumerate}
\item The Hamming distance counting not in common edges/non-edges between two graphs.
\item The symmetric difference between the cycles formed in two
graphs, i.e. counting the number of not in common cycles in two graphs.
\end{enumerate}

Hence, a mathematical representation of the constructed distance metric for two graphs $\mathcal{G}_{k},\mathcal{G}_{l}$ for $k,l\in \{1,\ldots,N\}$ is, 
\begin{displaymath}
d_{\text{HS}}(\mathcal{G}_{k},\mathcal{G}_{l})=d_{\text{H}}(A_{\mathcal{G}_{k}},A_{\mathcal{G}_{l}})+\lambda \cdot \mid C_{\mathcal{G}_{k}} \Delta C_{\mathcal{G}_{l}} \mid ,
\end{displaymath}

\noindent where $d_{\text{H}}(\cdot,\cdot)$ denotes the Hamming distance, $C_{\mathcal{G}_{i}}$ denotes the \am {directed} cycles in graph i, $\Delta$ indicates the symmetric difference and $\lambda \in \mathbb{R}$ is a weighting factor. In Supplementary material Section 1, we show that HS is a distance metric. Under this construction, we encode information about dissimilarities in the structure of the networks, with respect to both their edges and cycles. The tuning of the $\lambda$ parameter corresponds to how much influence we allow the symmetric difference to have on the total distance. 
In the rest of this article, we assume $\lambda$ to be equal to 1, suggesting equal importance between the Hamming and the Symmetric difference distance.

The specification of the HS distance metric for the SNF model induces significant challenges when adopting the MCMC framework proposed by \cite{lunagomez2020modeling} to make posterior inferences with the SNF model. \am{Notably the mixing of chains is very poor with acceptance rates close to zero, as illustrated in Figure \ref{is_trac2} in Section \ref{sec53}.} 
This motivated us to develop an alternative computational framework to make posterior inferences with the SNF model and details are given in Section \ref{sec:meth}.

\section{Overview of the SNF model}\label{sec:background}

In this section we provide a brief overview of the SNF model proposed in \cite{lunagomez2020modeling}, and why it is a compelling model to consider for this setting. We also highlight some shortcomings with the current implementation which limits its usefulness in our setting.

\subsection{Motivation and Model formulation}

\cite{lunagomez2020modeling} develop a model for network data inspired by the form of a Normal distribution. Specifically, they assume an underlying mean network representing the network population and a dispersion parameter denoting the variation of the networks about this mean. They express the mean network in terms of a Fr\'{e}chet mean, as seen in the studies of \cite{ginestet2017hypothesis} and \cite{kolaczyk2020averages}, and the dispersion parameter in terms of an entropy. Under this construction, they obtain the probabilistic mechanisms that generate data sets of multiple network data which they denote the SNF model. 

Specifically, if we assume we have a population of \am{directed} and unweighted graphs $\mathcal{G}_{1},\cdots,\mathcal{G}_{N}$ then the joint distribution characterised by the SNF model is given by,
\begin{equation}\label{snf}
P(A_{\mathcal{G}_{1}},\cdots,A_{\mathcal{G}_{N}}\mid A_{\mathcal{G}^{m}},\gamma)=\frac{1}{Z(A_{\mathcal{G}^{m}},\gamma)^N} \exp\left\{ -\gamma \cdot \sum_{i=1}^{N}\phi(d_{\mathcal{G}} (A_{\mathcal{G}_{i}},A_{\mathcal{G}^{m}}))\right\},
\end{equation}
where $\mathcal{G}^{m}$ is the Frech{\'e}t mean, $d_{\mathcal{G}}(\cdot,\cdot)$ is a distance metric, $ \gamma > 0$ is the dispersion, 
$\phi(\cdot)>0$ is a monotone increasing function and the model partition function is
\begin{displaymath}
Z(A_{\mathcal{G}^{m}},\gamma)=\sum_{A_{\mathcal{G}}\in \{ \mathcal{G}_{|n|} \}}\exp \{-\gamma \cdot \phi(d_{\mathcal{G}}(A_{\mathcal{G}},A_{\mathcal{G}^{m}}))\},
\end{displaymath}
where $\{\mathcal{G}_{|n|}\}$ is the space of $n$-node networks. The parameters $\mathcal{G}^{m}$, and $\gamma$ can thus be seen to relate to the mean and precision parameters in a Normal distribution.

\cite{lunagomez2020modeling} show the Centered Erd\"{o}s-R\'{e}nyi (CER) model is a special case of a SNF model when the Hamming distance metric is used. Under the CER model a population of networks is generated by perturbing the edges of a centroid network $\mathcal{G}^{m}$ using a Bernoulli distribution with probability $\alpha$, as follows:
\begin{equation}\label{samplingcer}
A_{\mathcal{G}}(i,j)\mid(A_{\mathcal{G}^{m}}(i,j),\alpha) = \mid A_{\mathcal{G}^{m}}(i,j)-Z(i,j)\mid,
\end{equation}
where $\mathcal{G}^{m}$ is the Frech{\'e}t mean and $Z(i,j)$'s are \emph{iid} $\mathrm{Ber}(\alpha)$, with $0< \alpha<0.5$. The joint distribution of a population of \am{directed} and unweighted $\mathcal{G}_{1},\cdots,\mathcal{G}_{N}$ graphs is then 
\begin{displaymath}
P(A_{\mathcal{G}_{1}},\cdots ,A_{\mathcal{G}_{N}}\mid A_{\mathcal{G}^{m}},\alpha)=\prod_{i=1}^{N}\alpha^{d_{H}(A_{\mathcal{G}_{i}},A_{\mathcal{G}^{m}})}\cdot(1-\alpha)^{n(n-1)-d_{H}(A_{\mathcal{G}_{i}},A_{\mathcal{G}^{m}})}
\end{displaymath}
where $d_{H}(\cdot,\cdot)$ denotes the Hamming distance metric and n is the number of nodes.

To make inferences, the authors adopt a Bayesian approach. A prior distribution for $\gamma$ is specified with support on $\mathbb{R}^{+}$. The prior choice and support is strongly related to the specified distance metric. A prior distribution for the network representative $A_{\mathcal{G}^{m}}$ is specified with the same functional form as that of the SNF model. The priors for the parameters of the CER model are specified in a similar manner, with the prior for the representative having the functional form of the CER model. The prior for $\alpha$ in the CER model requires support on $(0,0.5)$ with a scaled Beta distribution on $(0,0.5)$ proposed.

\subsection{Addressing the intractable normalising constant}

\cite{lunagomez2020modeling} make posterior inferences using a MCMC scheme to draw samples from the posterior distribution based on a Metropolis-Hastings (MH) algorithm. However, the normalising constant of the SNF model, $Z(A_{\mathcal{G}^{m}},\gamma)$, depends on the parameters of the model. Thus, the normalising constants do not cancel in the Metropolis-Hastings ratio.

To tackle the intractable normalising constants, \cite{lunagomez2020modeling} apply the Auxiliary Variable technique presented in \cite{moller2006}. Notably, \cite{moller2006} consider a likelihood of the form,
\begin{gather}\label{intrlik}
P(y\mid \theta)=\frac{q_{\theta}(y)}{\boldsymbol{Z}(\theta)},
\end{gather}
where $\theta$ denotes the model parameter, $y$ represents the data, $q_{\theta}(y)$ the unnormalised density, and $\boldsymbol{Z}(\theta)$ is an intractable normalising constant that depends on $\theta$. 

They propose the use of an auxiliary variable $x$ that has the same support as that of $y$, with density $f(x\mid\theta,y)$ to obtain an unbiased estimator of $\boldsymbol{Z}(\theta)$. In light of Importance Sampling, under \cite{moller2006} $\boldsymbol{Z}(\theta)$ can be written as,
\begin{equation}
\boldsymbol{Z}(\theta)=\mathbb{E}\bigg[\frac{q(x\mid\theta)}{f(x\mid\theta,y)}\bigg],
\end{equation}
where the expectation is taken with respect to the density of the auxiliary variable $x$, $f(x\mid\theta,y)$. In this regard, they propose sampling $x$ from $P(\cdot\mid\theta)$ \am{as seen in equation (\ref{intrlik})} and use the approximation,
\begin{equation}
\boldsymbol{Z}(\theta) \approx \frac{q(x\mid\theta)}{f(x\mid\theta,y)}.
\end{equation}
Thus, they can substitute the normalising constant $\boldsymbol{Z}(\cdot)$ by its unbiased estimator $q(x\mid\cdot)/f(x\mid\cdot,y)$ in the MH acceptance ratio. \amrev{In this respect, $q(x\mid\cdot)$ and $f(x\mid\cdot,y)$ are evaluated in each MCMC iteration at the auxiliary variable $x$ drawn from the proposal $P(\cdot\mid\theta)$.}

The formulation of the Auxiliary Variable Method in the case of the SNF model involves the simulation of a set of auxiliary variables $\mathcal{G}^{*}$, defined on the same state space as the network data $\{\mathcal{G}_{i}\}_{i=1}^{N}$. \cite{lunagomez2020modeling} exploit the probabilistic mechanism of the CER model to specify the conditional density $f(A_{\mathcal{G}^{*}_{1}}, \ldots,A_{\mathcal{G}^{*}_{N}} \mid \{\mathcal{G}_{i}\}_{i=1}^{N},A_{\mathcal{G}^{m}}, \tilde{\alpha})$ of the auxiliary network variables $\mathcal{G}^{*}_{1}, \ldots,\mathcal{G}^{*}_{N}$. Thence, in each iteration of the MH algorithm a new state of both the model parameters and the auxiliary network variables will be proposed, with the latter sampled from a proposal distribution that has the same functional form as the likelihood. Under this formulation, the normalising constants cancel in the MH ratio. For a more detailed description of this MH algorithm see \cite{lunagomez2020modeling}.

A main challenge in implementing the Auxiliary Variable Method for the SNF model is the slow mixing of the chain for $\gamma$, \am{as seen in Figure \ref{is_trac2} in Section \ref{sec53}.} 
Notably, we see a very low acceptance rate and thus poor mixing. Depending on the distance metric choice, this issue is apparent even for small network sizes. 

The occurrence of this phenomenon can be attributed to the discrepancy between the likelihood, the SNF model, and the choice of auxiliary density, the CER model. Depending on the choice of the distance metric for the SNF model, this discrepancy can increase leading to a bad mixing or, in some cases, the chain not exploring the state space at all. 

The poor mixing makes this impractical to consider for our setting and motivates our development of an alternative strategy to approximate the normalising constant. The proposed approach greatly improves performance of the MCMC, allowing it to be applied to similar sizes of networks present in the ecological data set.

\section{Proposed Bayesian inference framework for the SNF model using Importance Sampling}
\label{sec:meth}

To overcome shortcomings of the Auxiliary Variable approach
we develop an alternative method to approximate the intractable normalising constant. Specifically, we formulate an Importance Sampling step within our MCMC equivalent to Ratio Importance Sampling \citep{chen1997monte}. We were motivated by \cite{vitelli2017} who also use Importance Sampling to make Bayesian inference from the Mallow's model \citep{mallows1957non}, a common model for analysing rank data with the same functional form as the SNF model. Frequentist inference may also be possible, with \cite{mardia1999complex} developing this for the Watson model that also has the same functional form as the SNF model.

A key difference between the Mallow's model for rank data and the SNF model is that the normalising constant in the latter involves both the representative network and the dispersion parameter, while for the Mallow's model the normalising constant depends only on the dispersion parameter, for right-invariant distance metrics considered in \cite{vitelli2017}. This allows an off-line approximation of the normalising constant through IS, using a pseudo-likelihood approximation of the target distribution. 

Graphs are more complex objects than rank data, due to diverse structures they exhibit such as the formation of communities and motifs, as well as other topological structures revealed by their spectral decomposition. For networks, the right-invariance property does not hold for the majority of distance functions, with different properties governing rank and network data. Thus,
approximating the normalising constant of the SNF model is a more challenging scenario. Unlike \cite{vitelli2017}, we formulate an Importance Sampler within our MCMC to give a good approximation to the normalising constant. 

\subsection{Formulation of IS step for the SNF model}\label{subsec:is_snf}

The normalising constant of the SNF model has the following form,

\begin{equation}
Z(A_{\mathcal{G}^{m}},\gamma)=\sum_{A_{\mathcal{G}}\in \{ \mathcal{G}_{|n|} \}}\exp \{-\gamma \cdot \phi(d_{\mathcal{G}}(A_{\mathcal{G}},A_{\mathcal{G}^{m}}))\},
\end{equation}
this involves computing a sum over the space of $n$-node graphs, $\{\mathcal{G}_{|n|}\}$. \amrev{Even for modest $n$, this sum is impractical to compute.} 
Instead, using ideas from Importance Sampling \citep{robert2013monte} we can rewrite the sum as
\begin{equation}
\begin{gathered}
\sum_{A_{\mathcal{G}}\in \{ \mathcal{G}_{|n|} \}}\exp \{-\gamma \cdot \phi(d_{\mathcal{G}}(A_{\mathcal{G}},A_{\mathcal{G}^{m}}))\}= \\ \sum_{A_{\mathcal{G}}\in \{ \mathcal{G}_{|n|} \}} \frac{\exp \{-\gamma \cdot \phi(d_{\mathcal{G}}(A_{\mathcal{G}},A_{\mathcal{G}^{m}}))\}}{g(A_{\mathcal{G}})} g(A_{\mathcal{G}})= \\ \mathbb{E}_{g}\bigg[\frac{\exp \{-\gamma \cdot \phi(d_{\mathcal{G}}(A_{\mathcal{G}},A_{\mathcal{G}^{m}}))\}}{g(A_{\mathcal{G}})}\bigg],
\end{gathered}
\end{equation}
which can then be approximated by drawing a sample of networks $\mathcal{G}_{1},\ldots,\mathcal{G}_{K}$ from an Importance Sampling (IS) proposal density $g$ and calculating,
\begin{equation}\label{estz1}
\hat{Z}(A_{\mathcal{G}^{m}},\gamma)\approx \frac{1}{K} \sum_{k=1}^{K} \frac{\exp \{-\gamma \cdot \phi(d_{\mathcal{G}}(A_{\mathcal{G}_{k}},A_{\mathcal{G}^{m}}))\}}{g(A_{\mathcal{G}_{k}})}.
\end{equation}

One advantage of the IS method is the flexibility with specifying the IS density. In this regard, choices of distributions that are easy to sample from are preferred \citep{robert2013monte}. In our problem, a natural choice of the IS density is the distance-based CER model for two main reasons, (i) the CER model is a member of the Spherical Network Family (SNF) of models \citep{lunagomez2020modeling}, and (ii) sampling network data from the CER model is quick, thus will result in a less computationally intensive MCMC algorithm. 
\am{To sample networks from the CER model, we perturb the edges of the centroid $\tilde{A}_{\mathcal{G}^{m}}$ using Bernoulli noise with probability $\tilde{\alpha}$, as per equation (\ref{samplingcer}).}

Thus, the estimator in (\ref{estz1}) takes the following form under the CER IS density:

\begin{equation}\label{estz}
\hat{Z}(A_{\mathcal{G}^{m}},\gamma)\approx \frac{1}{K} \sum_{k=1}^{K} \frac{\exp \{-\gamma \cdot \phi(d_{\mathcal{G}}(A_{\mathcal{G}_{k}},A_{\mathcal{G}^{m}}))\}}{\tilde{\alpha}^{d_{\text{H}}(A_{\mathcal{G}_{k}},\tilde{A}_{\mathcal{G}^{m}})}(1-\tilde{\alpha})^{n(n-1)-d_{\text{H}}(A_{\mathcal{G}_{k}},\tilde{A}_{\mathcal{G}^{m}})}},
\end{equation}
where $\{A_{\mathcal{G}_{k}}\}_{k=1}^{K}$ are networks sampled from the CER model with parameters $\tilde{\alpha}$ and $\tilde{A}_{\mathcal{G}^{m}}$. 

We determine $\tilde{\alpha}$ and $\tilde{A}_{\mathcal{G}^{m}}$ by fitting the data to the CER model to obtain the posterior mean of $\tilde{\alpha}$ and posterior mode of $\tilde{A}_{\mathcal{G}^{m}}$. In this way, we encode information about the data that may allow a better approximation of the normalising constant. 

\subsection{MCMC scheme with IS step}\label{subsec:mcmc_is}

We now describe our computational framework to obtain posterior draws for the SNF model parameters. As seen in \cite{lunagomez2020modeling}, the joint posterior distribution of the centroid $A_{\mathcal{G}^{m}}$ and the dispersion parameter $\gamma$ can be expressed as

\begin{equation}\label{targ}
\begin{gathered}
P(\textcolor{black}{A_{\mathcal{G}^{m}}},\textcolor{black}{\gamma} \mid A_{\mathcal{G}_{1}},\cdots,A_{\mathcal{G}_{N}}) \propto \frac{1}{Z(A_{\mathcal{G}_{0}},\gamma_{0})}\exp \left\{-\gamma_{0}\phi(d_{\mathcal{G}}(A_{\mathcal{G}^{m}},A_{\mathcal{G}_{0}}))\right\} P(\gamma \mid \alpha_{0}) \cdot \\ \frac{1}{Z(A_{\mathcal{G}^{m}},\gamma)^N}\exp \{-\gamma \sum_{i=1}^{N} \phi(d_{\mathcal{G}}(A_{\mathcal{G}_{i}},A_{\mathcal{G}^{m}})) \}.
\end{gathered}
\end{equation}

We follow a largely similar scheme to \cite{lunagomez2020modeling} to make inferences, using Metropolis-Hastings to sample from the joint posterior of the parameters. However, to overcome the double-intractability problem, we approximate the normalising constant within each iteration of the MCMC using the estimator obtained through Importance Sampling, different to the Auxiliary Variable Method adopted by \cite{lunagomez2020modeling}. Notably, we obtain posterior draws from the target distribution in Equation (\ref{targ}), after substituting the normalising constant in the likelihood with its estimate in equation (\ref{estz}).

To obtain posterior draws for the parameters $A_{\mathcal{G}^{m}}$ and $\gamma$, we follow a similar scheme to \cite{lunagomez2020modeling}. 
Details are given in Supplementary material Section 2.

Algorithm \ref{algo1} sketched below illustrates the MH algorithm with IS step.


\begin{center}
\scalebox{0.8}{
\centering
\IncMargin{1em}
\RestyleAlgo{boxruled}
\begin{minipage}{1.1\linewidth}
\begin{algorithm}[H]
\KwData{$A_{\mathcal{G}_1},\ldots,A_{\mathcal{G}_N}$ \quad \textbf{Hyperparameters:} $A_{\mathcal{G}_0},\gamma_0,\alpha_0,\tilde{\alpha},\tilde{A}_{\mathcal{G}^m}$}
 
\BlankLine

\textbf{Initialisation:} 
Randomly generate $\gamma^{(0)}$ and $A_{\mathcal{G}^{m}}^{(0)}\sim\text{Bernoulli}(\sum_{i=1}^N A_{\mathcal{G}_i}/N)$ 
\BlankLine




\BlankLine
\For{$i\leftarrow 1$ \KwTo M}{

\textbf{MH step with a mixture of kernels:} Update $A_{\mathcal{G}^{m}}$ or $\gamma$
\newline 
Sample $v\sim \text{Multinomial}(\xi_{1},\ldots,\xi_{L})$
\newline
Depending on the value of $v$ propose $A_{\mathcal{G}^{m}}^{(i)}\sim q(A_{\mathcal{G}^{m}}^{(i)}\mid A_{\mathcal{G}^{m}}^{(i-1)})$ \newline
or $\gamma^{(i)}\sim q(\gamma^{(i)}\mid\gamma^{(i-1)})$ 


\textbf{Draw new IS sample of networks:} $A_{\mathcal{G}_1}^{IS(i)},\ldots,A_{\mathcal{G}_K}^{IS(i)}\sim \text{CER}(\tilde{\alpha},\tilde{A}_{\mathcal{G}^m})$.
\BlankLine

\BlankLine

\textbf{Estimate Z:} Use equation (\ref{estz}) to estimate normalising constant in posterior $P(A_{\mathcal{G}^{m}}^{(\cdot)},\gamma^{(\cdot)}\mid A_{\mathcal{G}_1},\ldots,A_{\mathcal{G}_N})$
\BlankLine
\textbf{Calculate MH ratio:}
r=$min\bigg(1,\frac{P(A_{\mathcal{G}^{m}}^{(i)},\gamma^{(i)}\mid A_{\mathcal{G}_1},\ldots,A_{\mathcal{G}_N})\cdot q(A_{\mathcal{G}^{m}}^{(i-1)},\gamma^{(i-1)}\mid A_{\mathcal{G}^{m}}^{(i)},\gamma^{(i)})}{P(A_{\mathcal{G}^{m}}^{(i-1)},\gamma^{(i-1)}\mid A_{\mathcal{G}_1},\ldots,A_{\mathcal{G}_N})\cdot q(A_{\mathcal{G}^{m}}^{(i)},\gamma^{(i)}\mid A_{\mathcal{G}^{m}}^{(i-1)},\gamma^{(i-1)})}\bigg)$

$u\sim\text{Bernoulli}(r)$

\eIf{u=1}
{Accept proposals $A_{\mathcal{G}^{m}}^{(i)},\gamma^{(i)}$}
{Reject proposals $A_{\mathcal{G}^{m}}^{(i)},\gamma^{(i)}$}
}

\caption{Metropolis-Hastings Algorithm with IS step}
\label{algo1}
\end{algorithm}
\end{minipage}
}
\end{center}

\begin{figure}[htb!]
    \centering
    \includegraphics[scale=.17]{ 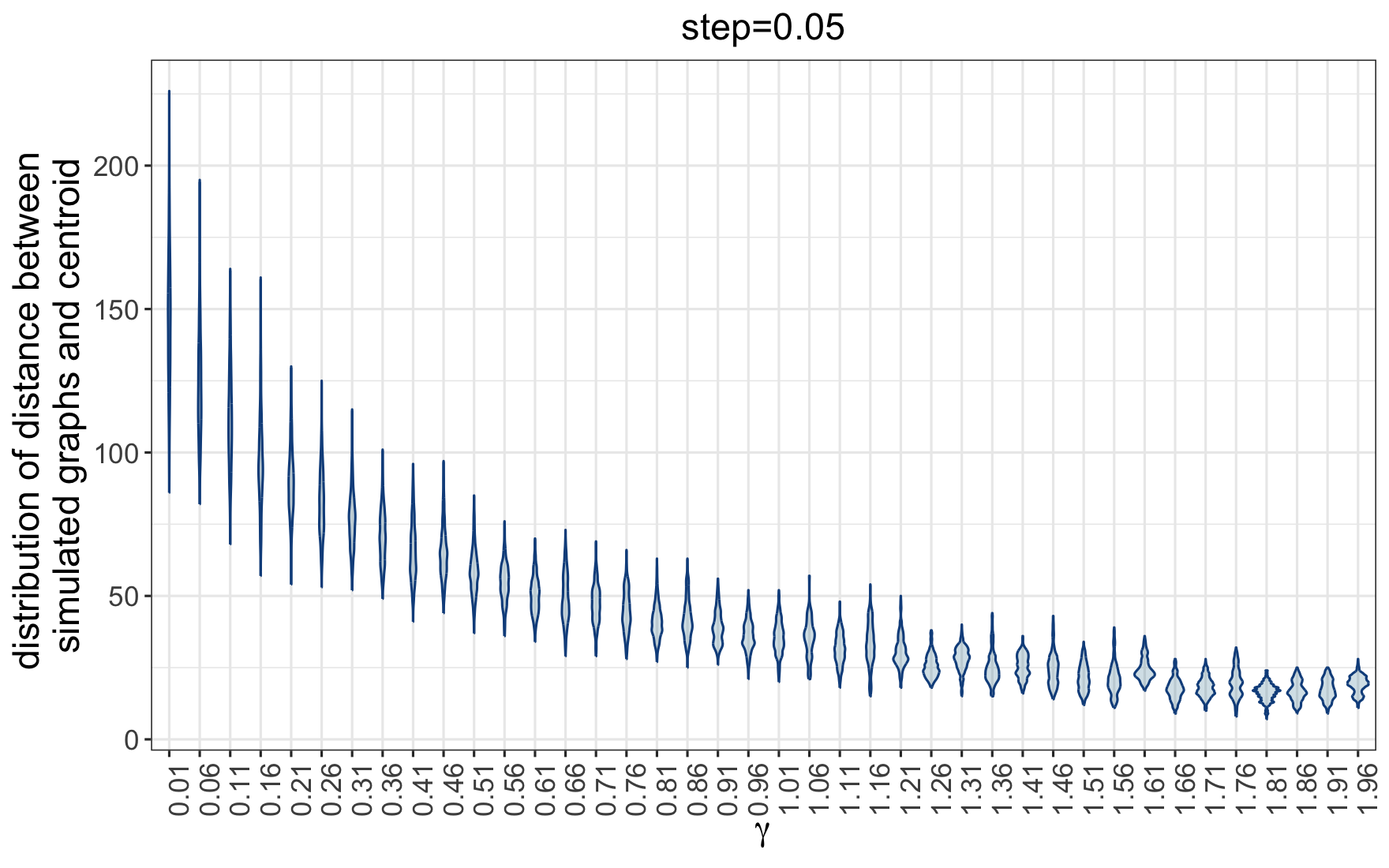}
    \caption{Distribution of HS distance between simulated graphs from SNF model and centroid $\mathcal{G}^m$, for varying $\gamma$ values.}
    \label{eda13}
\end{figure}

\am{We randomly generate an initial centroid network $\mathcal{G}^{m(0)}$ by sampling edges independently from a Bernoulli distribution with probabilities equal to the average adjacency matrix obtained from the observed network population $\sum_{i=1}^N A_{\mathcal{G}_i}/N$. Thus, we assist our MCMC with a meaningful network initialisation using information from the observed network population. In our implementation, we initialise $\gamma^{(0)}$ at 0.1. In practice, any real positive can be specified, ideally upper bounded by the value of $\gamma$ for which the average distance of the networks from the centroid is close to 0 (see Figure \ref{eda13}). For the mixture of kernels we consider proposals imposing both moderate and more drastic changes in the current values of the parameters. Our investigation suggests that inferences are not sensitive to moderate changes in the probabilities. The prior specification for the centroid and the dispersion is performed as suggested in \cite{lunagomez2020modeling}.}

\am{\subsection{Addressing the MCMC chain mixing issue}\label{sec53}}

\am{In this section, we illustrate the improvement in the MCMC chain mixing using the IS step compared to the auxiliary variable technique used in \cite{lunagomez2020modeling}, for the HS distance metric.}

\begin{figure}[htb!]
    \centering
    \includegraphics[scale=.15]{ 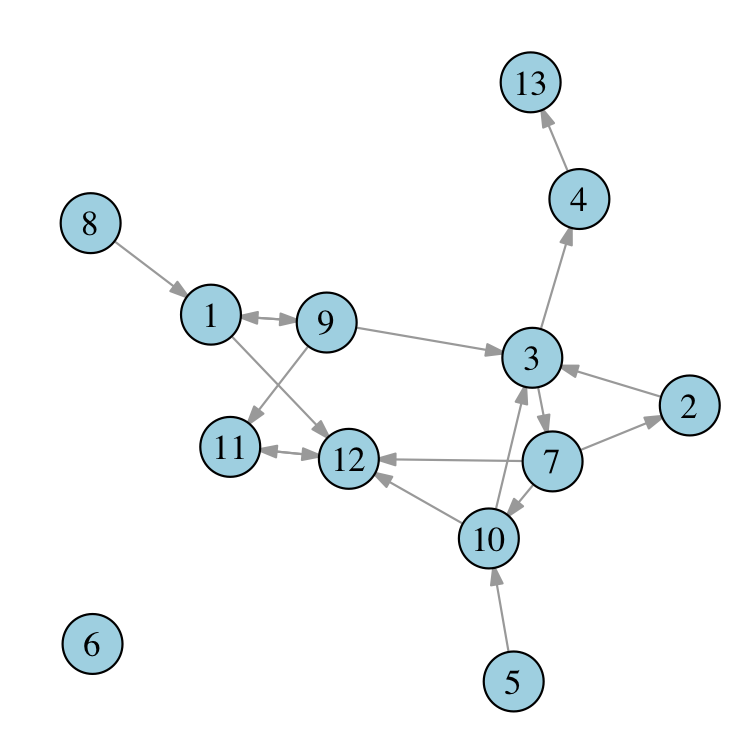}
    \caption{Simulated centroid with 13 nodes.}
    \label{simcentr13node}
\end{figure}

\begin{figure*}[htb!]
\centering
\begin{minipage}[b]{0.48\linewidth}
\centering
\includegraphics[height=2in]{ 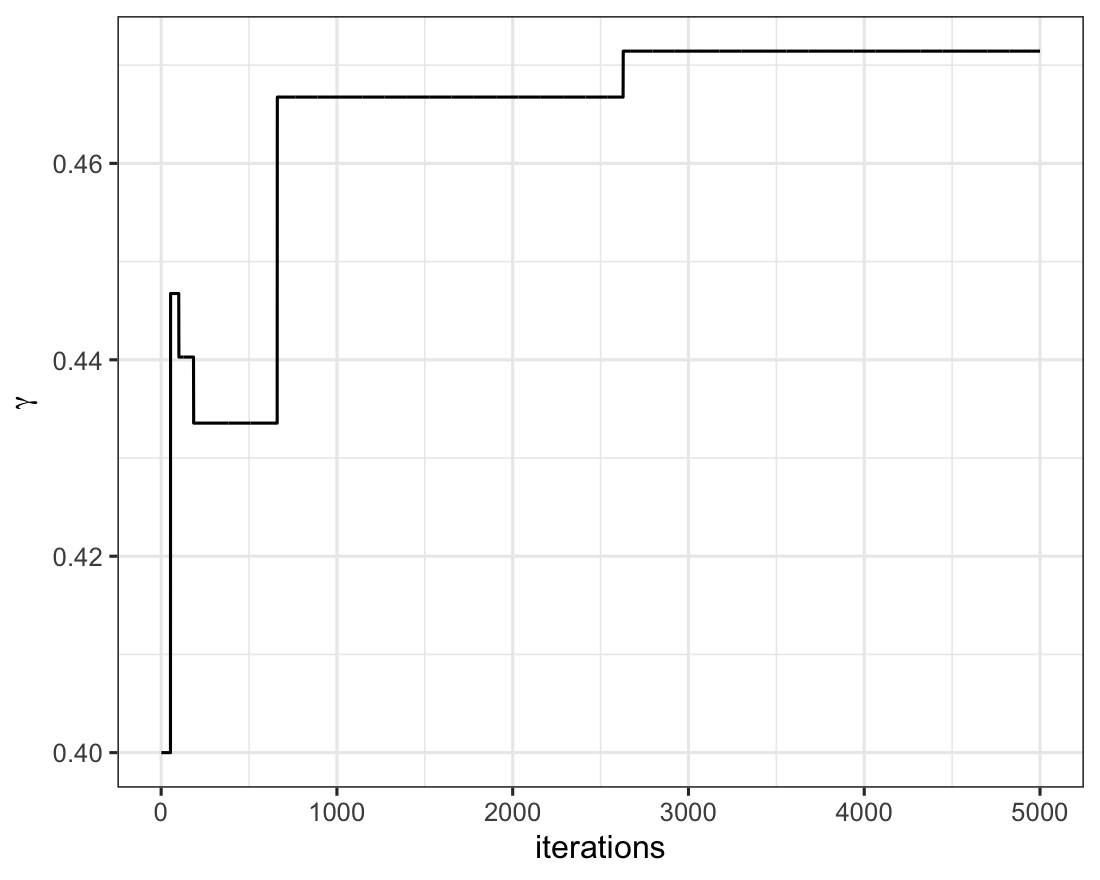}
\end{minipage}
\quad
\begin{minipage}[b]{0.48\linewidth}
\centering
\includegraphics[height=2in]{ 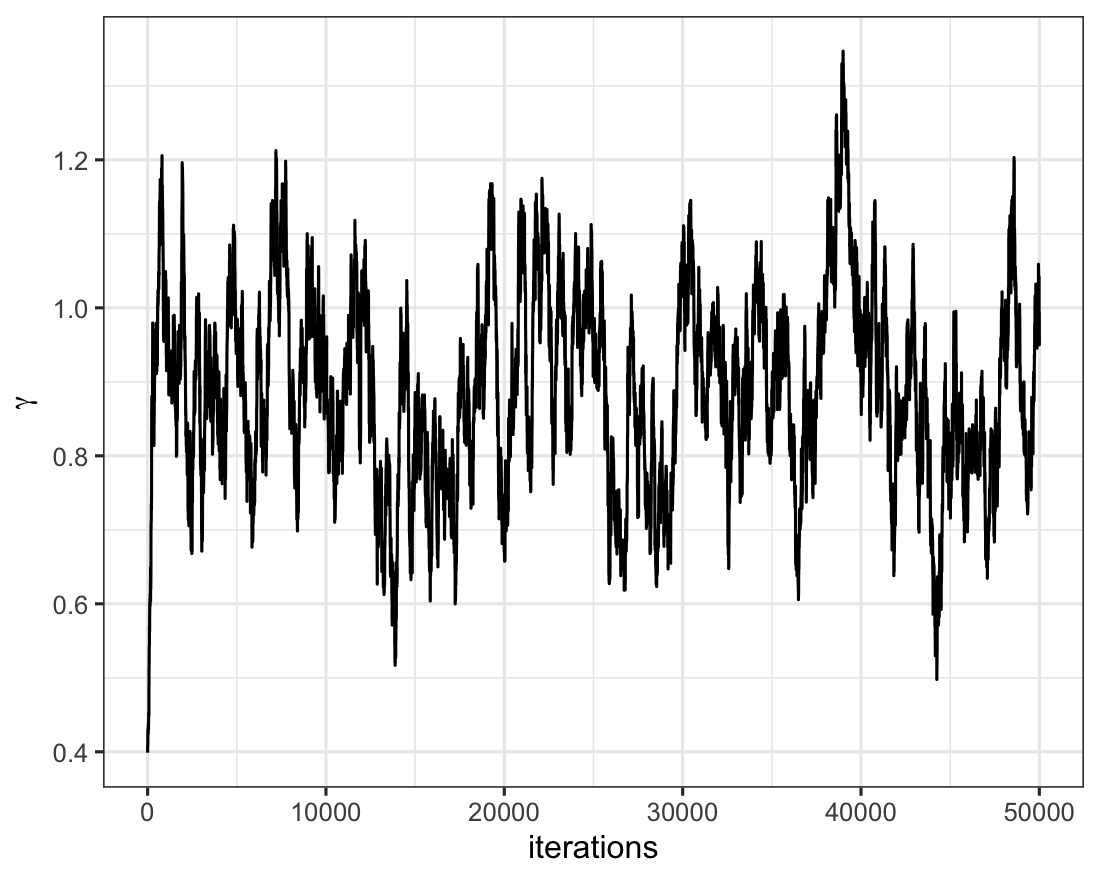}
\end{minipage}
\caption{MCMC samples for the dispersion parameter $\gamma$ in the SNF model with HS distance metric using the auxiliary variable method in \cite{lunagomez2020modeling} (left), versus using our framework utilising Importance Sampling (right), for the same simulated network population.}\label{is_trac2}
\end{figure*}

\am{In this simulation experiment, we consider network size, population size and parameter values similar to that of the ecological application. In particular, we simulate a population of $N=13$ networks with $n=13$ nodes, under the scenario of a 13-node centroid with density approximately 0.1 (Figure \ref{simcentr13node}) and $\gamma=1.2$. Similarly to \cite{lunagomez2020modeling}, we simulate from the SNF model with this parameter specification using an MH algorithm with target distribution the density of the SNF model as seen in equation (\ref{snf}).}
\am{In Figure \ref{is_trac2}, we show the results after running the MCMC using the auxiliary variable technique for 5,000 iterations (left) and our proposed MCMC with IS step for 50,000 iterations (right), with an IS sample of 3000 networks. We observe a drastic improvement in the mixing of the MCMC chain using our proposed MCMC scheme.}

\am{The auxiliary variable technique requires sampling from the SNF model using an MH algorithm in each MCMC iteration, resulting in a computationally intensive algorithm. Indicatively, running the MCMC with the auxiliary variable technique for 5,000 iterations in this simulation experiment requires approximately 20 hours. In contrast, our proposed MCMC scheme with IS step does not require running an MH algorithm in each MCMC iteration since we can sample directly from the CER model using (\ref{samplingcer}). In this simulation experiment, running our proposed MCMC with IS step for 50,000 iterations requires approximately 20 hours. Running each MCMC scheme for 5,000 iterations and 50,000 iterations respectively, results in running the two approaches for a comparable amount of time. Thus, our proposed approach not only improves mixing but also substantially improves computation time.}

\am{We observe that the posterior region for $\gamma$  explored by the MCMC is slightly lower than the true value of $\gamma=1.2$ specified to simulate the network population. An explanation is possible by examining the EDA violin plots obtained in Figure \ref{eda13}, showing the distribution of the HS distance between the centroid and simulated networks from the SNF model for different $\gamma$ values. We see when the true value of $\gamma$ lies in $(0.9,1.21)$ the distribution of the distance between the simulated networks and the centroid is similar. Thus, for this regime, changes in the parameter space result in very small changes in the distribution, making the estimation of $\gamma$ in this regime a more challenging task.  
}

\amrev{The scope of this experiment is to illustrate the performance of our method and highlight the improvement of the MCMC chain mixing compared to the auxiliary variable technique, in a similar setting to that of the real data application.} \textcolor{black}{In the next section, we consider a regime with higher noise and a centroid with more cycles, to further illustrate the performance of our method in recovering $\gamma$ and the practicality of our proposed approach in recovering cycles compared to the baseline CER model.}

\subsection{\amrev{Network population with more cycles and high noise}}

\amrev{We consider a simulation regime with a 13-node centroid enclosing 26 directed triangles, and high noise indicated by a smaller size of $\gamma$. Notably we specify $\gamma=0.2$ for which regime the distribution of distance in its neighbourhood is more distinguishable. We simulate a population of $N=100$ networks using the SNF model with HS distance.}

\begin{figure*}[htb!]
\centering
\begin{minipage}[b]{0.4\linewidth}
\centering
\includegraphics[height=1.9in]{ 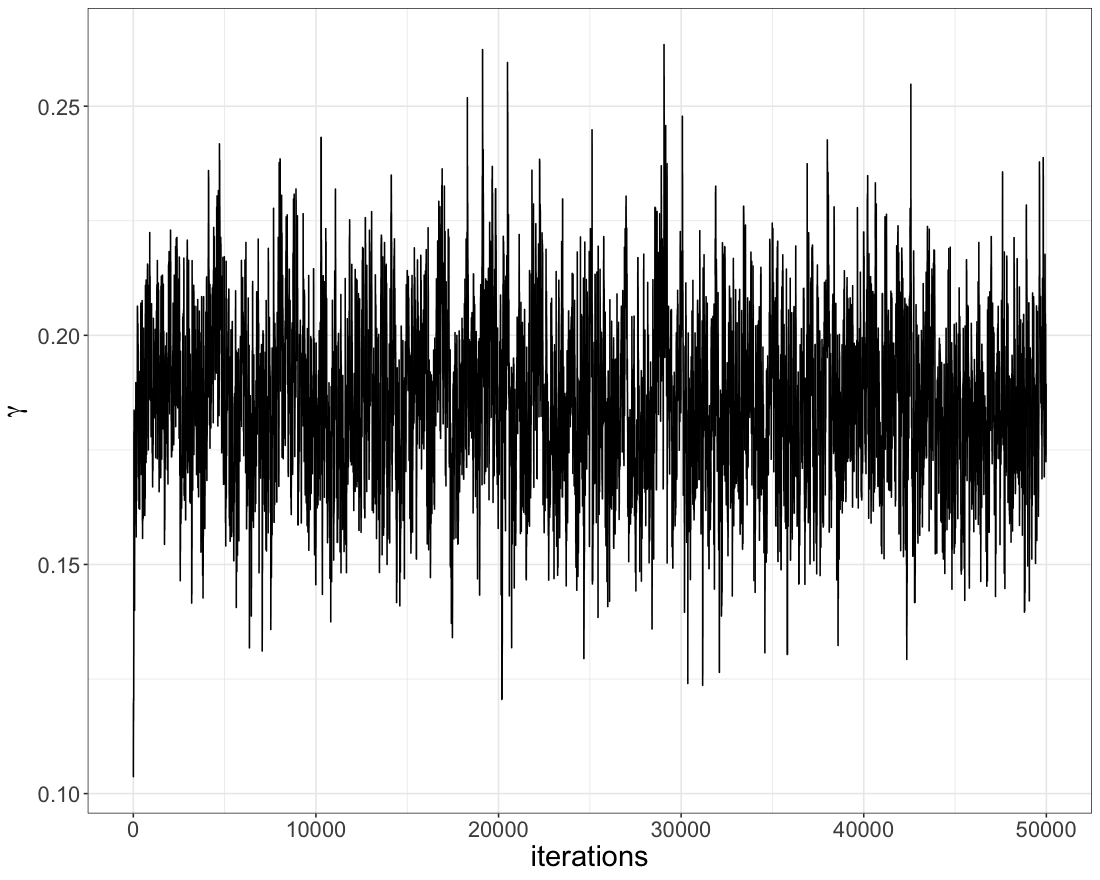}
\end{minipage}
\quad
\begin{minipage}[b]{0.56\linewidth}
\centering
\includegraphics[height=1.9in]{ 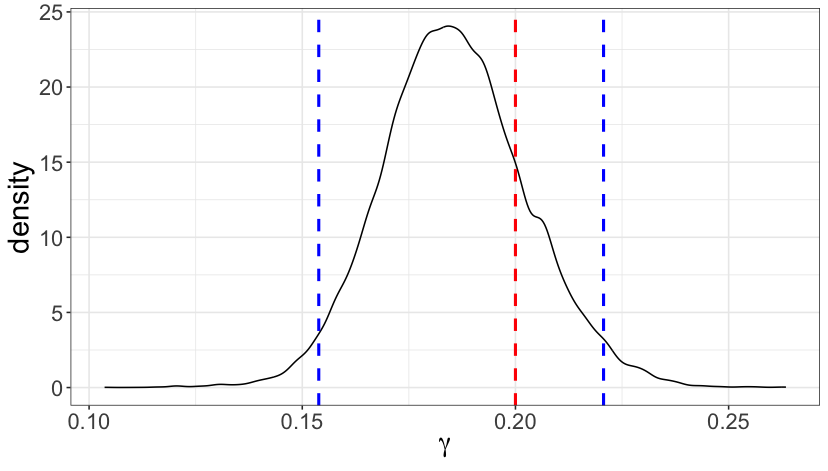}
\end{minipage}
\caption{MCMC samples for the dispersion parameter $\gamma$ in the SNF model with HS distance metric using IS (left), histogram of MCMC samples with blue dashed lines indicating the 95\% credible interval and red dashed line indicating the true size of $\gamma=0.2$ (right).}\label{sim_N100}
\end{figure*}

\amrev{Figure \ref{sim_N100} shows the traceplot of the posterior draws for $\gamma$ for 50,000 iterations of the MCMC, and the histogram of the posterior distribution for $\gamma$ with the 95\% credible interval (blue dashed lines) and the true size of $\gamma$ (red dashed line). The results suggest accurate recovery of the true $\gamma$ with posterior mean 0.19 and the true value of $\gamma$ lying within the 95\% credible interval. }

\begin{figure}[htb!]
    \centering
    \includegraphics[height=2.2in]{ 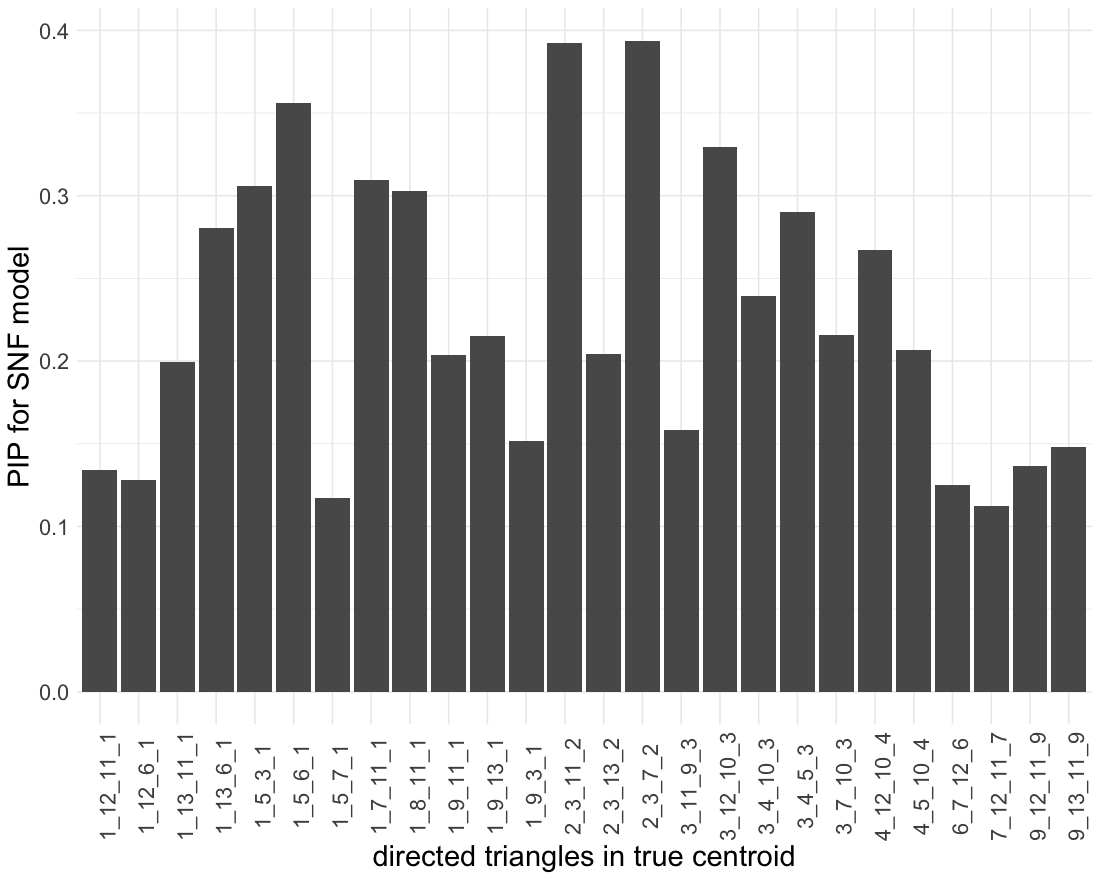}
  \\
    \includegraphics[height=2.2in]{ 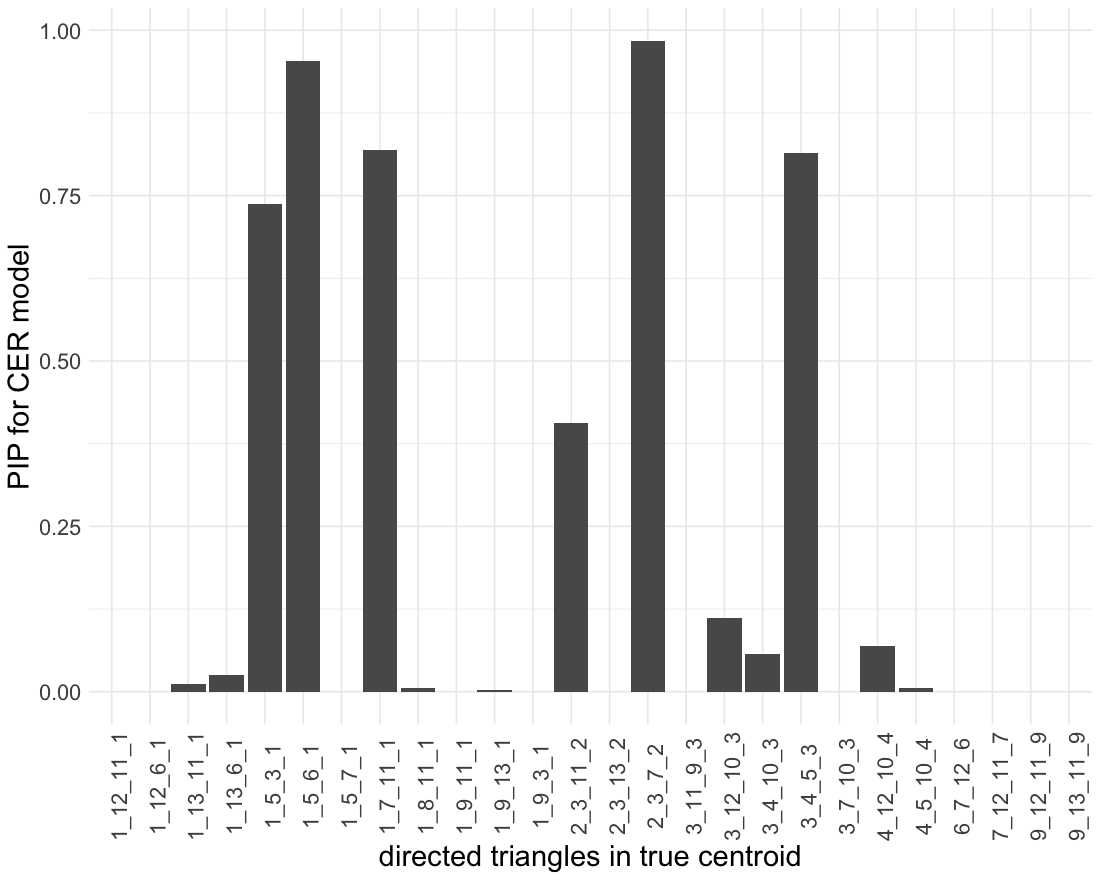}
    \caption{Posterior Inclusion Probabilities (PIP) for directed triangles in true centroid for MCMC samples of SNF model with HS distance metric using IS (top), and of CER model (bottom).}
    \label{pip}
\end{figure}

\amrev{We also examine the ability of our approach in recovering the cycles of the true centroid and compare it to the CER model as a baseline. We summarise the results from the posterior draws for the centroid by calculating the Posterior Inclusion Probability (PIP) of the directed triangles which are present in the true centroid, for the SNF with HS distance and the CER model respectively, as shown in Figure \ref{pip}. The results shown are for 50,000 MCMC iterations with a burn-in of 1,000 iterations for the SNF model with HS distance, and for 200,000 MCMC iterations with a burn-in of 150,000 iterations for the CER model. We observe that the SNF model with HS distance is able to recover all the directed triangles present in the true centroid with PIP up to 0.4, while the CER model is able to recover only approximately half of the true directed triangles (14 out of 26) with up to 0.98 PIP. This result highlights the key difference between the two models; on the one hand the SNF model with HS distance explicitly accounts for directed cycles but with smaller PIP due to the presence of high noise with respect to directed triangles in the simulated population. On the other hand the directed triangles detected using the CER model are only incidental and an artifact of the frequency of specific edges within the simulated network population.}

\am{\subsection{Empirical evaluation of normalising constant approximation}}

\textcolor{black}{We now evaluate the performance of our IS approximation to the normalising constant in a simulation setting.
Specifically, we consider the approximation applied to 4-node directed networks. Here there are $2^{12}=4,096$ possible networks, which is a small enough set to calculate the normalising constant exactly under a given model parameterisation in each iteration of the MCMC. This is then used as a benchmark with which to compare the normalising constant approximated through IS.} 

\begin{figure}[htb!]
    \centering
    \includegraphics[scale=.1]{ 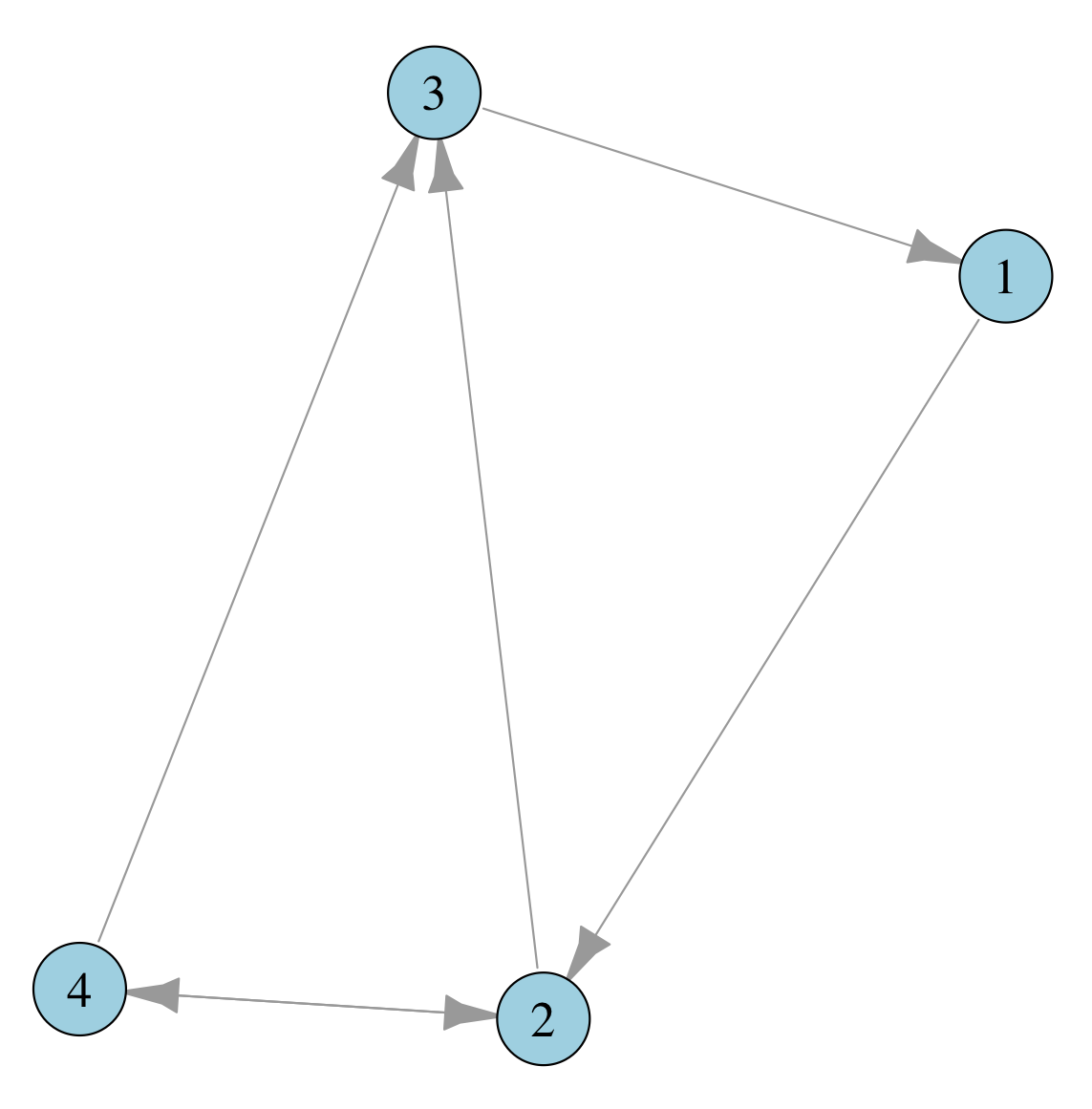}
    \caption{Simulated centroid with 4 nodes.}
    \label{4nodecentr}
\end{figure}

\begin{figure}[htb!]
\centering
\includegraphics[height=2.2in]{ 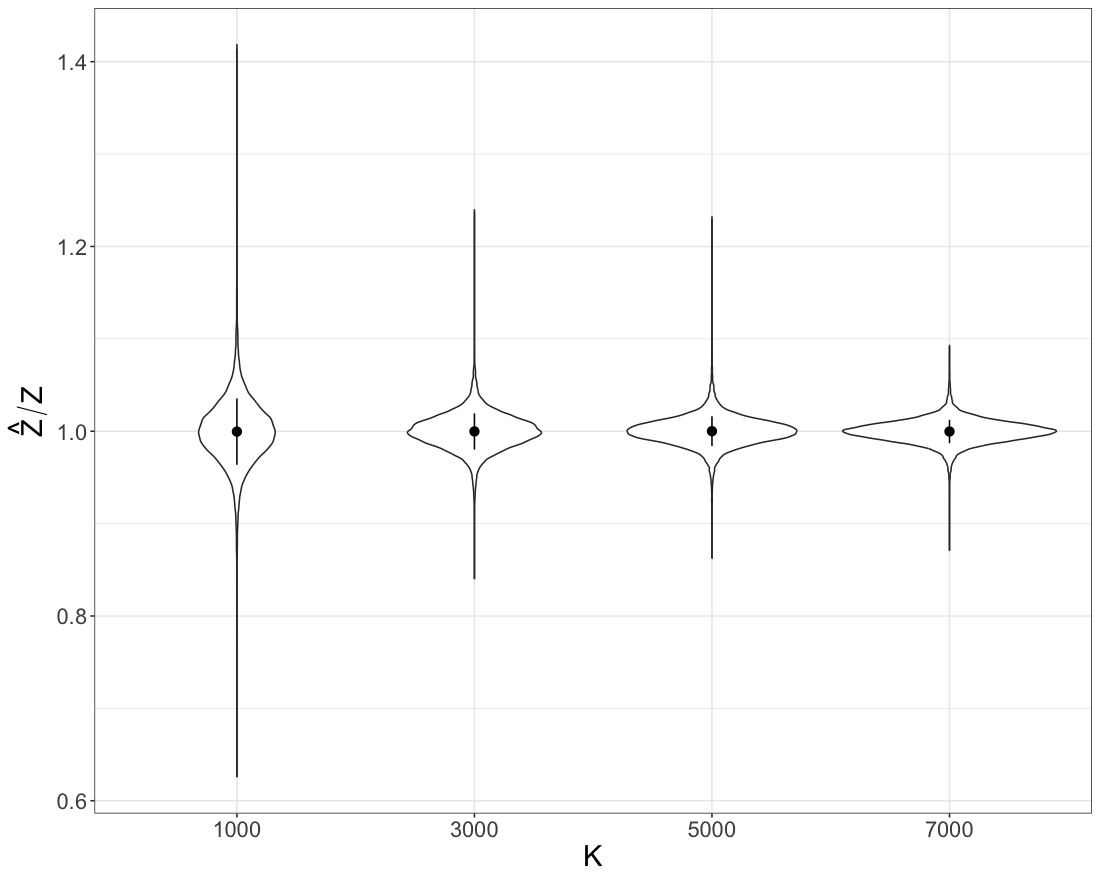}
\caption{Distribution of ratio $\hat{Z}(A_{\mathcal{G}^{m}},\gamma)/Z(A_{\mathcal{G}^{m}},\gamma)$ for 50,000 iterations of our MCMC scheme with IS step \amrev{for $K=\{1000,3000,5000,7000\}$}, for 4-node directed networks.}\label{ztrue}
\end{figure}

\am{We simulate a population of $N=13$ networks, similarly to the population size in our ecological application, using the SNF model with dispersion $\gamma=1$ and 4-node directed network centroid, as shown in Figure \ref{4nodecentr}, enclosing 1 directed triangle. We run our MCMC scheme with IS step for 50,000 iterations, 
and in each iteration
, we also calculate the true normalising constant. \amrev{To further explore the sensitivity of the approximation to the IS sample size $K$, we consider a range of values for $K=\{1000,3000,5000,7000\}$}. In Figure \ref{ztrue} we present the 
\amrev{distribution} of the ratio of the estimated normalising constant $\hat{Z}(A_{\mathcal{G}^{m}},\gamma)$ and the exact normalising constant $Z(A_{\mathcal{G}^{m}},\gamma)$ \amrev{for varying $K$ sizes and} 50,000 iterations of our MCMC. 
\amrev{We observe that the distribution of the ratio has mean (points in Figure \ref{ztrue}) and median equal to 1 for all $K$, and standard deviation (error bars in Figure \ref{ztrue}) of 0.036, 0.019, 0.016 and 0.013 for each $K$ respectively. The approximation is only marginally sensitive to the size of $K$ for $K\ge 3000$, which justifies the choice of an IS sample $K=3000$ networks, to avoid additional computational complexity of cycle detection for large samples of networks in each MCMC iteration.
The results indicate that our proposed IS step not only improves the mixing of the MCMC chain (see Section \ref{sec53}), but it is also a good approximation with respect to the exact $Z(A_{\mathcal{G}^{m}},\gamma)$ even for small $K=1000$.}
}

\newpage
\am{\section{\davidsal{Reef fish aggressive interactions}}\label{sec:realdata}}

\am{We analyse data collected at various reefs at different regions (Philippines, Bali, Christmas, Iriomote) in the Indo-Pacific ocean \citep{keith2018synchronous}. Each network observation represents the competitive interactions between species of fish. \textcolor{black}{ 
We align the node set across all reefs to comprise the fish species that are in common across all regions, resulting in networks with 13 nodes. The node labels used to represent the species are presented in Table \ref{befspecies}. The resulting number of reefs (networks) in our sample is 13, 
and are visually represented in Supplementary material, Section 3.}} 
\am{In this network population, there are three reefs for which directed triangles are observed, as shown in Table \ref{cycnetpop}. We observe that there are no reefs sharing common cycles. }

\am{To fit the SNF model with HS distance, we first tune the prior distributions of the parameters and the IS density. We tune the prior for the centroid and dispersion parameter in a similar manner as in \cite{lunagomez2020modeling}. Specifically, we centre the prior for the centroid at the network observation that minimises the distance from the rest of the networks in the sample \citep{lunagomez2020modeling}. We specify a Gamma prior distribution for the dispersion parameter $\gamma$ and center it with respect to the average HS distance of the network data from the centroid estimate,}
\am{where the centroid estimate is obtained by majority vote (connect nodes $i$ and $j$ in centroid estimate if the majority of the network population has an edge between $i$ and $j$). The size of each IS sample is set to $K=3000$ networks.}

\begin{table}[htb!]
\centering
\begin{tabular}{lr}
  \hline
  species & label \\ 
  \hline
 auriga & 1 \\ 
  baronessa & 2 \\ 
 citrinellus & 3 \\ 
 ephippium & 4 \\ 
 kleinii & 5 \\ 
   lunula & 6 \\ 
  lunulatus & 7 \\ 
 ornatissimus & 8 \\ 
  rafflesii & 9 \\ 
 speculum & 10 \\ 
  trifascialis & 11 \\ 
 unimaculatus & 12 \\ 
 vagabundus & 13 \\ 
   \hline
\end{tabular}
\caption{Node labels for species across all regions.}\label{befspecies}
\end{table}

\begin{table}[htb!]
\centering
\begin{tabular}{ll}
\multicolumn{1}{c}{\textbf{Reef}} & \multicolumn{1}{c}{\textbf{Cycle}} \\ \hline
Jemeluk                           & 2-3-5-2                            \\
                                  & 2-5-7-2                            \\
                                  & 2-9-5-2                            \\
                                  & 3-5-7-3                            \\
                                  & 3-5-13-3                           \\
                                  & 5-7-9-5                            \\
Lipah                             & 2-11-3-2                           \\
Nata                              & 7-12-11-7                         
\end{tabular}
\caption{Triangles in observed networks (reefs).}\label{cycnetpop}
\end{table}

\begin{figure}[htb!]
    \centering
    \includegraphics[scale=.18]{ 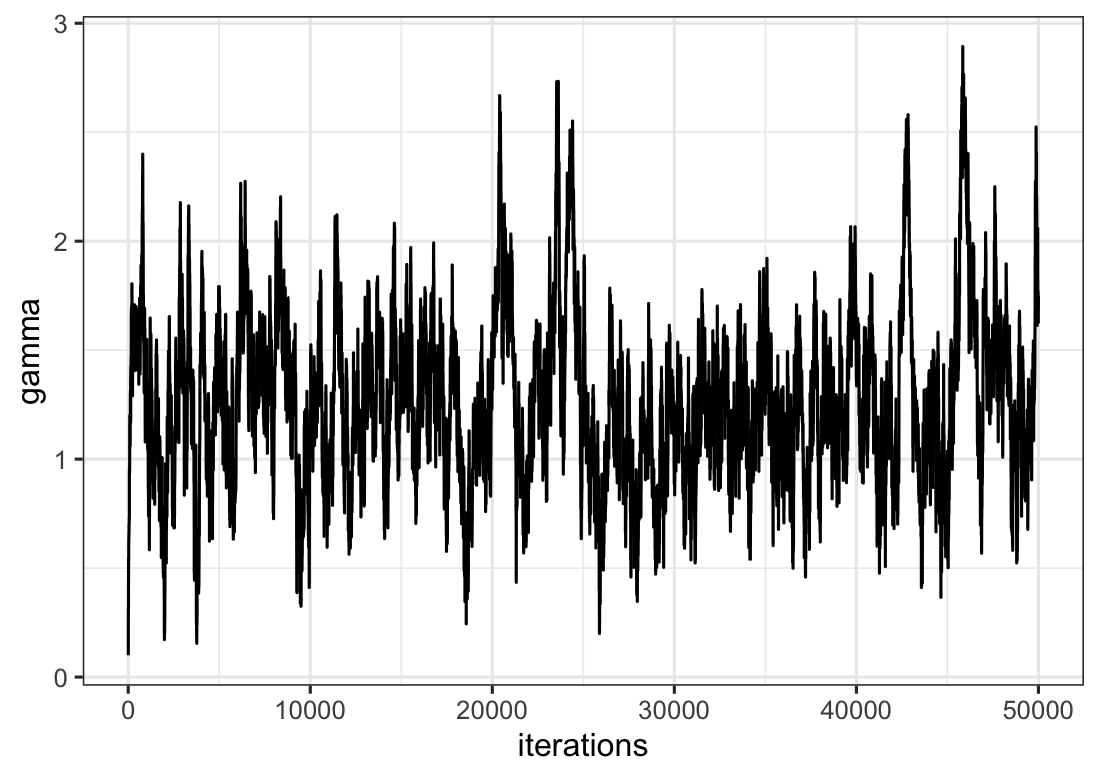}
    \caption{Traceplot for $\gamma$ for SNF fitted on fish networks from all regions.}
    \label{allreefsgamma2}
\end{figure}

\am{We run our MCMC for 50,000 iterations and obtain summaries of the posterior centroids and associated posterior draws of the dispersion parameter $\gamma$ of the SNF model. Figure \ref{allreefsgamma2} shows a traceplot for the parameter $\gamma$. }

\am{To investigate our model's efficacy in capturing directed cycle information in the network data, we obtain the 10 most common directed triangles enclosed in the posterior centroids, along with the proportion of the posterior centroid samples containing each cycle, and detect whether these cycles are also observed in the data. In Table \ref{cyclesbound} we present the 10 most common directed triangles identified in the posterior draws for the centroid network after 10,000 iterations burn-in, along with the proportion of times identified and whether the cycle is observed in the network data (observed) or not observed in the network data (inferred).}

\am{We observe that 4 out of the 10 directed triangles in the top 10 most common directed triangles are also \davidsal{observed in the real} data, with the rest of them enclosing nodes only from the set of nodes observed in the directed triangles present in the network population (Table \ref{cycnetpop}). This indicates that our MCMC algorithm meaningfully accepts posterior centroids with respect to directed triangles observed in the network population. \textcolor{black}{Moreover, there is evidence to suggest that the model is assigning posterior weight to directed triangles based on information in the network data as opposed to simply sampling networks with directed triangles formed by randomly picking 3 unique nodes from amongst those that form the observed directed triangles present in the data (comprising 8 distinct nodes). If it were we would expect each of the $8 \choose 3$ possible directed triangles here to have approximately equal posterior inclusion probabilities of 1/56 but the most commonly identified cycles have posterior inclusion probabilities much greater than this.} 
}

\begin{figure}[htb!]
\centering
\includegraphics[width=.4\textwidth]{ 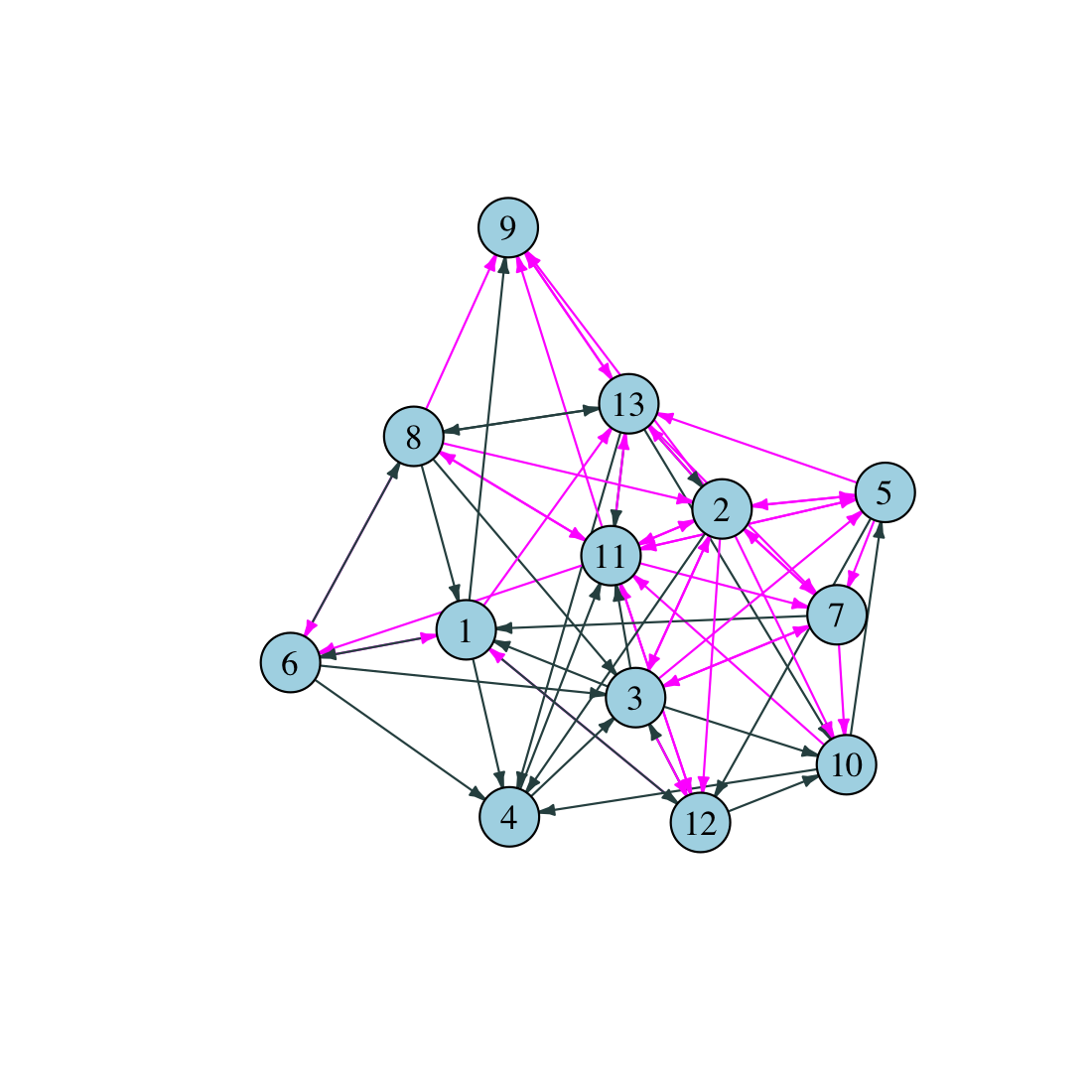}\hfill
\includegraphics[width=.4\textwidth]{ 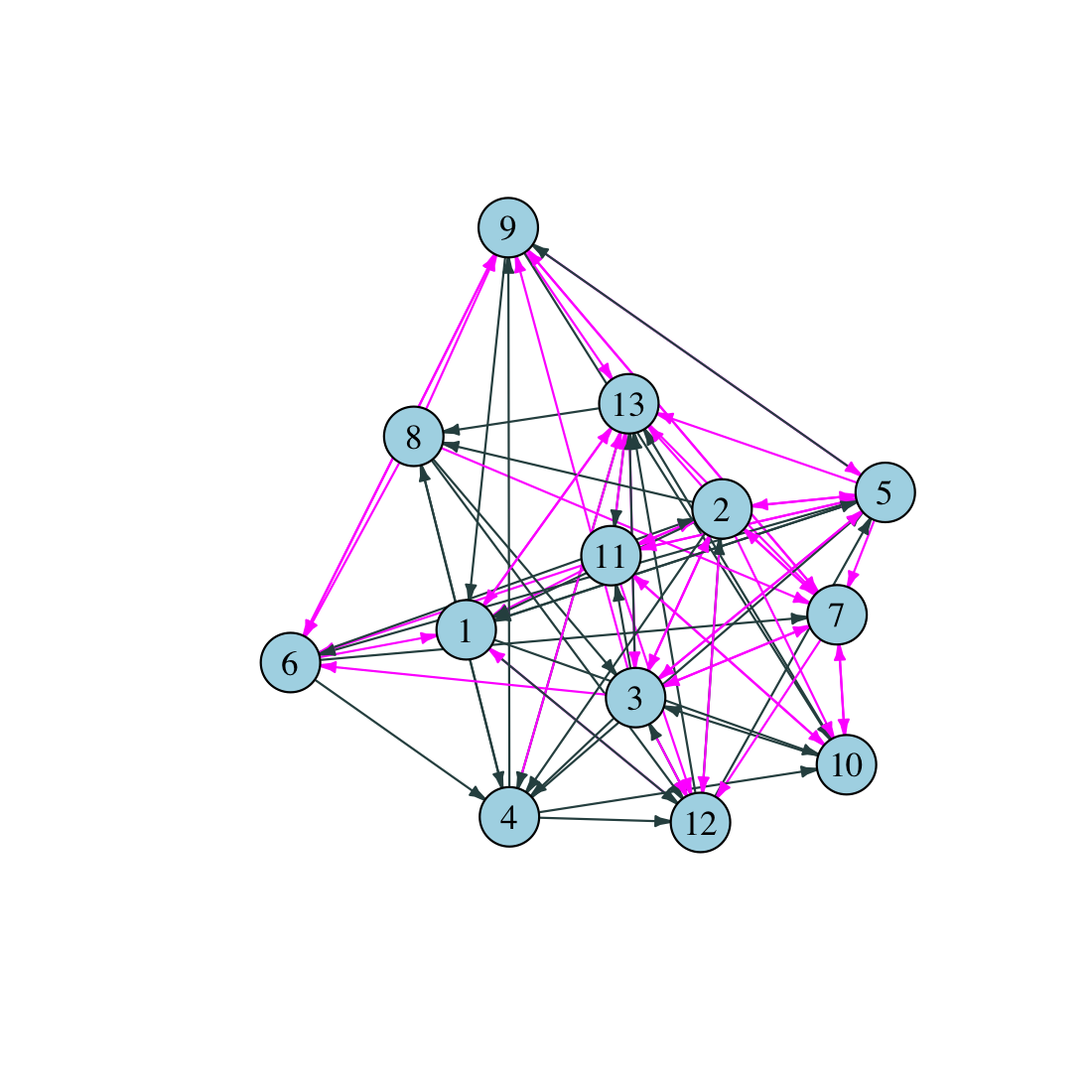}\hfill
\includegraphics[width=.3\textwidth]{ 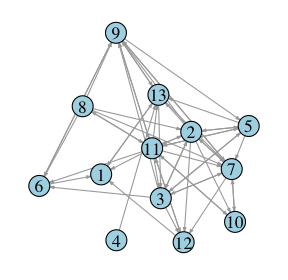}
\caption{Two posterior centroids with highest posterior mass 0.002 (top left) and 0.0015 (\amrev{top right}). Pink edges indicate edges also present in the data. Signal to noise ratio (number of edges in common/number of edges inferred) 1.44 and 1.38 respectively. \amrev{Union of network population (bottom).}}
\label{postmodes_allreefs50K}
\end{figure}

\am{In Figure \ref{postmodes_allreefs50K}, we further illustrate the two networks \amrev{(top graphs)} with highest posterior mass. We highlight which edges of these posterior samples are also present in \amrev{any of }the network data in pink. We \textcolor{black}{note} that the two posterior centroids, taken from the high posterior mass region, have small posterior mass. This is to be expected when making inferences across a large space of possible graphs coupled with a diverse set of network data. \amrev{The network at the bottom in Figure \ref{postmodes_allreefs50K} encloses the union of the edges of all observed networks for ease of comparison.}}

\begin{table}[htb!]
\centering
\begin{tabular}{ccc}
\hline
\multicolumn{3}{c}{\textbf{SNF model with HS distance directed triangles}} \\ \hline
\textbf{\begin{tabular}[c]{@{}c@{}}most common\\ directed triangles\end{tabular}} & \textbf{\begin{tabular}[c]{@{}c@{}}proportion of \\ times identified\end{tabular}} & \textbf{\begin{tabular}[c]{@{}c@{}} observed/ \\inferred\end{tabular}} \\ \hline 
2-3-7-2 & 0.52 & inferred \\ 
3-5-7-3 & 0.49 & observed \\ 
2-5-11-2 & 0.49 & inferred \\ 
2-5-7-2 & 0.48 & observed \\ 
2-5-3-2 & 0.48 & inferred \\ 
3-9-5-3 & 0.47 & inferred \\ 
2-9-7-2 & 0.46 & inferred \\ 
3-5-13-3 & 0.46 & observed \\ 
2-9-11-2 & 0.43 & inferred \\ 
2-3-5-2 & 0.42 & observed \\ \hline
\end{tabular}
\caption{Most common directed triangles in posterior draws for centroid $\mathcal{G}^m$.}\label{cyclesbound}
\end{table}

\am{For comparison we fit the CER model to the data, that only considers the Hamming distance. 
After running the CER model on the network population for 50,000 iterations with a burn-in of 10,000 iterations, we observe that the model \textcolor{black}{is unable} to make inferences on directed triangles. Notably, none of the posterior centroids encloses a directed triangle, despite the \textcolor{black}{presence of} directed triangles in the network population. This is anticipated for two reasons, (i) as the CER model assumes that the centroid is polluted by Bernoulli noise, increasing the network population results in Bernoulli noise corrupting the cycles in the data rather than preserving them, and (ii) transitivity for Erd\"{o}s-R\`{e}nyi models is very low. In contrast, the SNF model explicitly accounts for directed cycles through the HS distance metric. This finding highlights the importance of our proposed modelling framework when it is of interest 
to capture the formation of directed cycles in a network population.}

\section{Conclusion}
\label{sec:conc}

Modelling multiple network data is essential to addressing many applied research questions. In this article we proposed a metric that explicitly incorporates networks' cycles, denoted as the HS distance. We incorporated this metric within the SNF model and developed a computational framework to allow posterior inferences in practical ecological settings, hitherto not possible with the original implementation of the model. We applied our modelling framework to make inferences from ecological data studying aggressive interactions between species of fish and were able to infer cyclical properties that were not possible to detect using a simpler CER model that does not account for cycle information.

While we have shown benefits of our approach \textcolor{black}{in the analysis of} ecological studies, cycles are of interest in many other fields such as neuroscience or genetics. The HS distance can thus be an informative measure in many other network applications. A key challenge in these settings may be the larger size of networks (number of nodes). Dealing with cycles is already a substantial computational endeavour, and this issue will be exacerbated in larger networks. \textcolor{black}{Specifically,} \amrev{cycle detection is the main bottleneck for scaling to larger networks, since our approach requires the use of an IS sample of networks and the calculation of cycles within this sample.} 
\textcolor{black}{We addressed this here through restricting the model to only consider directed triangles. This was relevant for the ecological study which focused on intransitive competition, \davidsal{and triadic isomorphs are commonly analysed in ecology due to their ease of interpretation.}}
Considering other strategies to deal with the computational burden of calculating cycles would also be of interest. \textcolor{black}{ A possible direction would be to obtain an approximation of the HS distance metric using a machine learning approach.} 

\amrev{Another interesting direction for future research would be to consider a model selection framework for the SNF model, which could potentially inform the size of $\lambda$ in the HS distance metric for a given data example. }

The ecological study representing the species interactions also contained edges weights, corresponding to the multiple interactions observed between species of fish. It would be interesting to consider how this feature could be incorporated within our modelling framework. A modification of the Hamming distance in the HS metric would be required to quantify dissimilarities between weighted graphs, with the Frobenius distance an alternative. We would also need to adapt the computational framework to permit sampling of networks from a weighted space of graphs. This would be necessary for both sampling network representatives in the MCMC as well for approximating the normalising constant through Importance Sampling. There is no straightforward solution to addressing this, but if a model could be developed and implemented practically, it would open up the possibility to model a wide range of weighted networks previously not possible with existing methods. 

There have been various techniques developed to approximate intractable normalising constants. We implemented an IS that offered substantial advantages over the original Auxiliary Variable method proposed for the SNF model in \cite{lunagomez2020modeling}. It would be interesting to explore whether other approximations might also confer advantages and develop an appreciation for which methods are best suited for different settings.

\textcolor{black}{More generally, the flexibility of the SNF model offers the potential to address many applied research questions when used to analyse populations of network data. It would be interesting to consider whether the SNF model could be further developed to construct formal hypothesis tests that permit additional model based inferences in this setting. 
The mathematical challenges to overcome here are non-trivial and present an interesting avenue for future research.}

As our desire to analyse more complex data structures increases so do the modelling and computational challenges. Our methodology for incorporating network cycles for statistical modelling and inference of ecological data has opened up an exciting new area within analysis of multiple networks to explore. Accordingly, there is the potential to build on this and address a number of important questions in the field, both theoretical and applied.

\section*{Supplementary information}
The supplement to "Bayesian modelling and computation utilising directed cycles in multiple network data" contains additional details about our proposed distance metric, methodology and real data.




\textbf{Authors' contributions:} Conceptualization: Anastasia Mantziou, Robin Mitra, Sim\'{o}n Lunag\'{o}mez, Sally Keith; Methodology: Anastasia Mantziou, Robin Mitra, Sim\'{o}n Lunag\'{o}mez; Formal analysis and investigation: Anastasia Mantziou; Writing - original draft preparation: Anastasia Mantziou; Writing - review and editing: Robin Mitra, Sim\'{o}n Lunag\'{o}mez, Sally Keith, David Jacoby; Funding acquisition: Anastasia Mantziou; Resources: Sim\'{o}n Lunag\'{o}mez, Sally Keith; Supervision: Sim\'{o}n Lunag\'{o}mez, Robin Mitra.

\bibliography{bibsam}

\end{document}


\maketitle

\noindent%
{\it Keywords:} Doubly intractable distributions, Importance Sampling, Object data analysis, Relational data.
\vfill

\newpage

\section{Proof HS is a distance metric}

The HS measure is the weighted sum of the Hamming distance and the symmetric difference of cycles between two graphs. The Hamming distance is a well-known distance metric, thus, to prove that the HS measure is also a distance metric, we need to prove that the symmetric difference between graphs' cycles is a distance metric.

Let $\mathcal{C}_n$ be the set of cycles for graphs of size $n$, and each $C_{\mathcal{G}_i},C_{\mathcal{G}_j},C_{\mathcal{G}_k}\in \mathcal{C}_n$ be the subset of cycles found in graphs $\mathcal{G}_i$, $\mathcal{G}_j$ and $\mathcal{G}_k$ respectively. Thence, the symmetric difference of the cycles of two graphs is $d_{symm}=\mid C_{\mathcal{G_{\cdot}}}\Delta C_{\mathcal{G_{\cdot}}}\mid$. The function $d_{symm}: \mathcal{C}_n \times \mathcal{C}_n \rightarrow [0,\infty )$ is a distance metric if the following conditions are satisfied:
\begin{enumerate}
\item $d_{symm}(C_{\mathcal{G}_i},C_{\mathcal{G}_j})=0 \Leftrightarrow C_{\mathcal{G}_i}=C_{\mathcal{G}_j}$ 
\item $d_{symm}(C_{\mathcal{G}_i},C_{\mathcal{G}_j})=d_{symm}(C_{\mathcal{G}_j},C_{\mathcal{G}_i})$
\item $d_{symm}(C_{\mathcal{G}_i},C_{\mathcal{G}_j})\leq d_{symm}(C_{\mathcal{G}_i},C_{\mathcal{G}_k})+d_{symm}(C_{\mathcal{G}_k},C_{\mathcal{G}_j})$
\end{enumerate} 

Conditions 1 and 2 are clearly satisfied. Thus, we need to prove that the triangle inequality holds for the symmetric difference of cycles. The symmetric difference has the following property,

\begin{equation*}
C_{\mathcal{G}_i}\Delta C_{\mathcal{G}_j}=(C_{\mathcal{G}_i}\Delta C_{\mathcal{G}_k})\Delta (C_{\mathcal{G}_k}\Delta C_{\mathcal{G}_j}).
\end{equation*}
It follows that 
\begin{gather*}
C_{\mathcal{G}_i}\Delta C_{\mathcal{G}_j}\subseteq (C_{\mathcal{G}_i}\Delta C_{\mathcal{G}_k})\cup (C_{\mathcal{G}_k}\Delta C_{\mathcal{G}_j}) \Rightarrow \\
\mid C_{\mathcal{G}_i}\Delta C_{\mathcal{G}_j}\mid\leq \mid C_{\mathcal{G}_i}\Delta C_{\mathcal{G}_k}\mid +\mid C_{\mathcal{G}_k}\Delta C_{\mathcal{G}_j}\mid .
\end{gather*}
Thus condition 3 is satisfied for the symmetric difference of cycles between graphs.

\section{Additional details for the Proposed Bayesian inference framework for the SNF model using Importance Sampling}

We now present additional details on the inferential scheme used to obtain draws from the posterior distributions of the parameters of the SNF model, as discussed in Section 5.2 of the main article. Notably, we update the adjacency matrix of the centroid $A_{\mathcal{G}^{m}}$ using either of the following two proposals,

\begin{itemize}
\item[(I)] We perturb the edges of the current centroid $A_{\mathcal{G}^{m}}^{(curr)}$ as follows:
\begin{displaymath}
A_{\mathcal{G}^{m}}^{(prop)}(i,j)=
\begin{cases}
1-A_{\mathcal{G}^{m}}^{(curr)}(i,j), \text{ with probability } \omega \\
A_{\mathcal{G}^{m}}^{(curr)}(i,j), \text{ with probability } 1-\omega
\end{cases}.
\end{displaymath}
\item[(II)] We propose a new network representative $A_{\mathcal{G}^{m}}^{(prop)}$, with each edge of the proposed representative being drawn independently from a Bernoulli distribution with parameter $\frac{1}{N}\sum_{l=1}^{N}A_{\mathcal{G}_{l}}(i,j)$, where $\{A_{\mathcal{G}_{l}}\}_{l=1}^{N}$ denoting the $N$ observed networks.
\end{itemize} 

Under case (I), we accept the proposed network representative $A_{\mathcal{G}^{m}}^{(prop)}$ with probability
\begin{align*}
\min \Bigg\{ 1,\frac{ \widehat{Z}(A_{\mathcal{G}^{m}}^{(prop)},\gamma^{(curr)})^{-N}\exp \{-\gamma^{(curr)} \sum_{i=1}^{N} d_{\mathcal{G}}(A_{\mathcal{G}_{i}},A_{\mathcal{G}^{m}}^{(prop)}) \}}{ \widehat{Z}(A_{\mathcal{G}^{m}}^{(curr)},\gamma^{(curr)})^{-N} \exp \{-\gamma^{(curr)} \sum_{i=1}^{N} d_{\mathcal{G}}(A_{\mathcal{G}_{i}},A_{\mathcal{G}^{m}}^{(curr)}) \}}\cdot  \\ \frac{\exp \left\{-\gamma_{0}d_{\mathcal{G}}(A_{\mathcal{G}^{m}}^{(prop)},A_{\mathcal{G}_{0}})\right\}}{\exp \left\{-\gamma_{0}d_{\mathcal{G}}(A_{\mathcal{G}^{m}}^{(curr)},A_{\mathcal{G}_{0}})\right\}} \Bigg\},
\end{align*}
while under case (II), we accept the proposed network representative $A_{\mathcal{G}^{m}}^{(prop)}$ with probability
\begin{align*}
\min \Bigg\{ 1,\frac{ \frac{\exp \{-\gamma^{(curr)} \sum_{i=1}^{N} d_{\mathcal{G}}(A_{\mathcal{G}_{i}},A_{\mathcal{G}^{m}}^{(prop)}) \}}{\widehat{Z}(A_{\mathcal{G}^{m}}^{(prop)},\gamma^{(curr)})^{N}}\exp \left\{-\gamma_{0}d_{\mathcal{G}}(A_{\mathcal{G}^{m}}^{(prop)},A_{\mathcal{G}_{0}})\right\}}{ \frac{\exp \{-\gamma^{(curr)} \sum_{i=1}^{N} d_{\mathcal{G}}(A_{\mathcal{G}_{i}},A_{\mathcal{G}^{m}}^{(curr)}) \}}{\widehat{Z}(A_{\mathcal{G}^{m}}^{(curr)},\gamma^{(curr)})^{N}} \exp \left\{-\gamma_{0}d_{\mathcal{G}}(A_{\mathcal{G}^{m}}^{(curr)},A_{\mathcal{G}_{0}})\right\}}\cdot  \\ \frac{Q(A_{\mathcal{G}^{m}}^{(curr)}\mid A_{\mathcal{G}^{m}}^{(prop)})}{Q(A_{\mathcal{G}^{m}}^{(prop)}\mid A_{\mathcal{G}^{m}}^{(curr)})} \Bigg\},
\end{align*}
We note here that the proposal distribution under case (I) is symmetric, and thus it cancels out from the Metropolis ratio, while under case (II) the proposal distribution $Q(A_{\mathcal{G}^{m}}^{(\cdot)}\mid A_{\mathcal{G}^{m}}^{(\cdot)})$ does not cancel.

Accordingly, we use a mixture of $K$ random walks to propose values for the dispersion parameter $\gamma$, as follows:
\begin{enumerate}
\item Draw a uniform random variable $u\sim \text{Unif}(-v_{k},v_{k})$, with k indicating the $k^{th}$ proposal.
\item Perturb the current state $\gamma^{(curr)}$ by the uniform random variable drawn, \\ $y=\gamma^{(curr)}+u$.
\item The newly proposed value for $\gamma$ is 
$\gamma^{(prop)}=\begin{cases}y, \text{if } y>0 \\ -y, \text{if } y<0 \end{cases}$,
\end{enumerate}
which we accept with probability
\begin{align*}
\min \Bigg\{ 1,\frac{ \widehat{Z}(A_{\mathcal{G}^{m}}^{(curr)},\gamma^{(prop)})^{-N} \exp \{-\gamma^{(prop)} \sum_{i=1}^{N} d_{\mathcal{G}}(A_{\mathcal{G}_{i}},A_{\mathcal{G}^{m}}^{(curr)}) \}}{\widehat{Z}(A_{\mathcal{G}^{m}}^{(curr)},\gamma^{(curr)})^{-N}\exp \{-\gamma^{(curr)} \sum_{i=1}^{N} d_{\mathcal{G}}(A_{\mathcal{G}_{i}},A_{\mathcal{G}^{m}}^{(curr)}) \}} \cdot \\ \frac{P(\gamma^{(prop)} \mid \alpha_{0})}{P(\gamma^{(curr)} \mid \alpha_{0})} \Bigg\}.
\end{align*}
Under this scheme, in each iteration of the MCMC algorithm, we draw a new sample from the IS density to calculate $\widehat{Z}$ in the numerator and denominator of the MH ratio, as detailed in Sections 5.1 and 5.2 of the main article. 

\section{Additional details for real data application}
\vspace{-2cm}
\begin{figure}[htb!]
\centering
\begin{minipage}[b]{0.3\linewidth}
\centering
\includegraphics[height=1.5in]{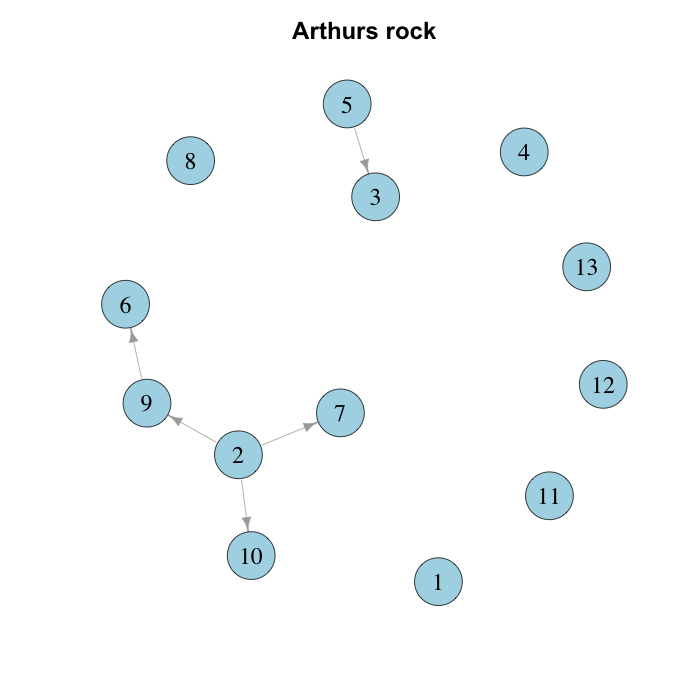}
\end{minipage}
\quad
\begin{minipage}[b]{0.3\linewidth}
\centering
\includegraphics[height=1.5in]{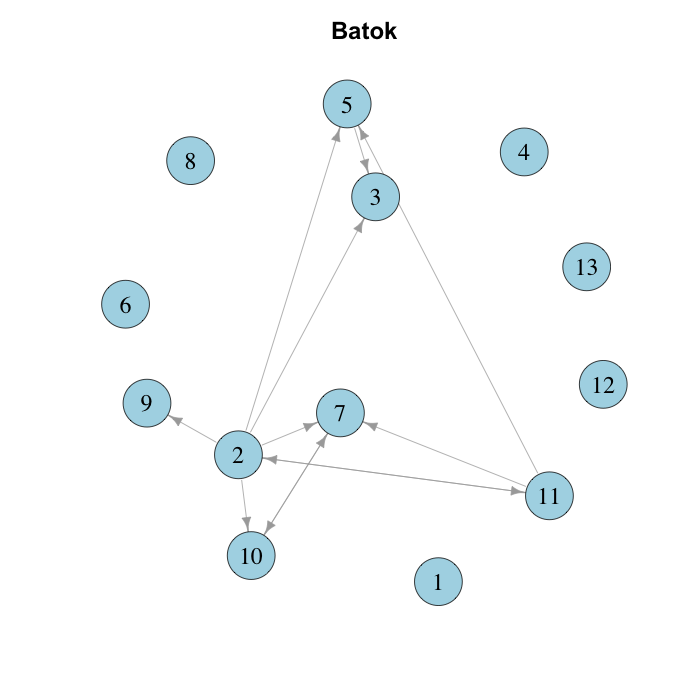}
\end{minipage}
\begin{minipage}[b]{0.3\linewidth}
\centering
\includegraphics[height=1.5in]{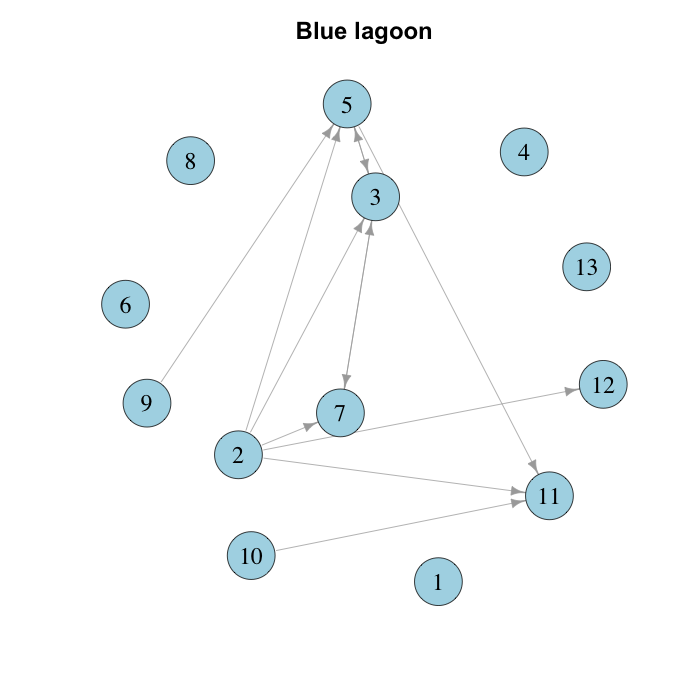}
\end{minipage}
\hfill
\begin{minipage}[b]{0.3\linewidth}
\centering
\includegraphics[height=1.5in]{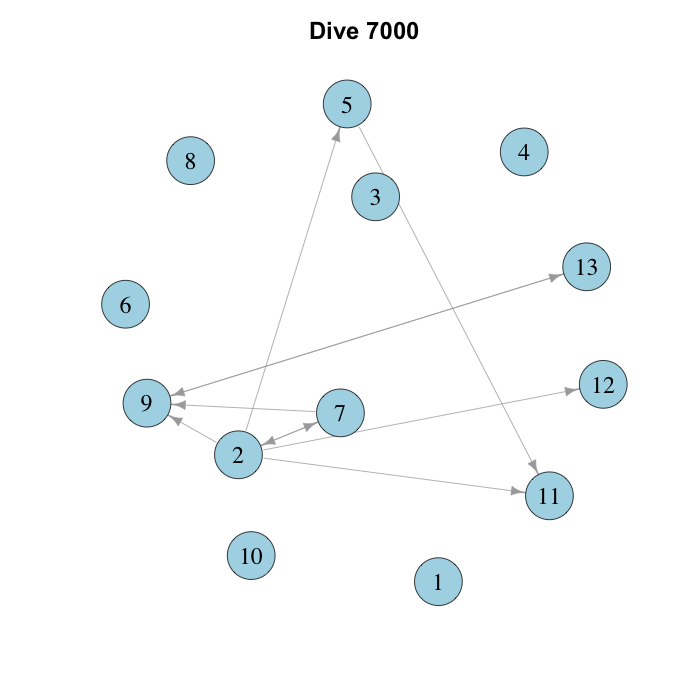}
\end{minipage}
\begin{minipage}[b]{0.3\linewidth}
\centering
\includegraphics[height=1.5in]{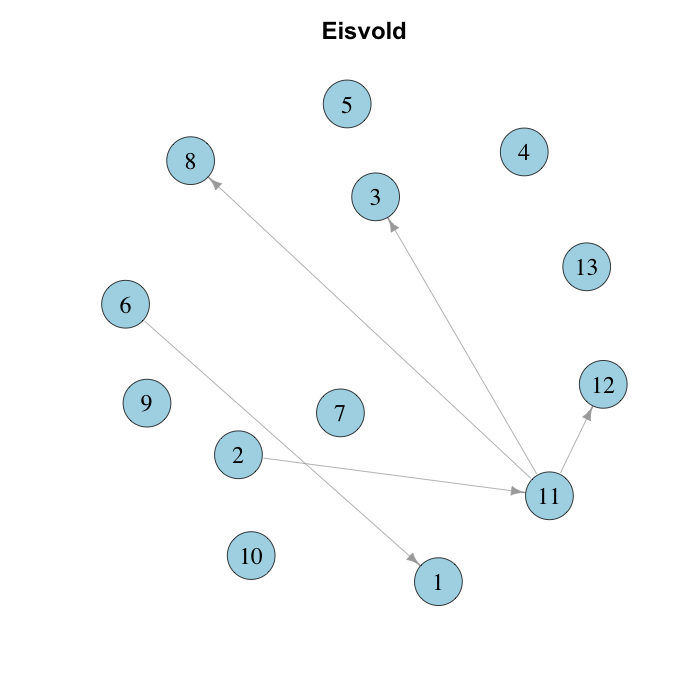}
\end{minipage}
\begin{minipage}[b]{0.3\linewidth}
\centering
\includegraphics[height=1.5in]{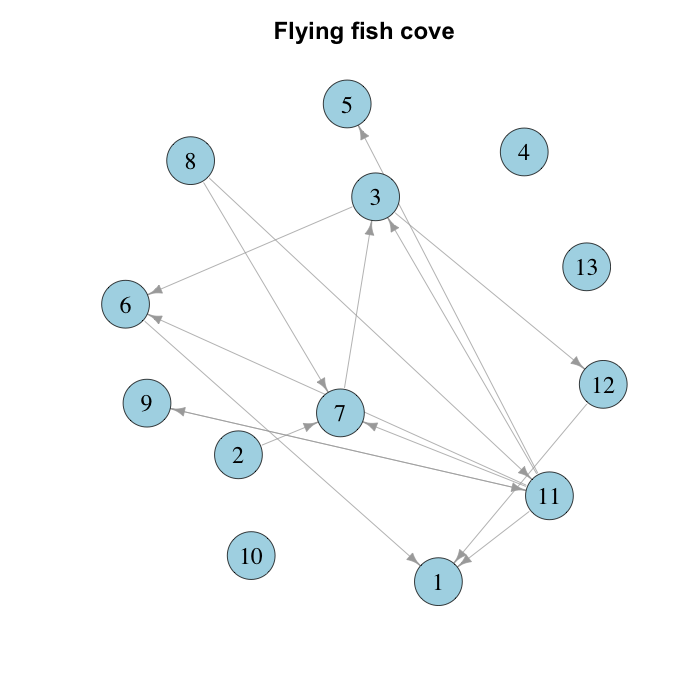}
\end{minipage}
\hfill
\begin{minipage}[b]{0.3\linewidth}
\centering
\includegraphics[height=1.5in]{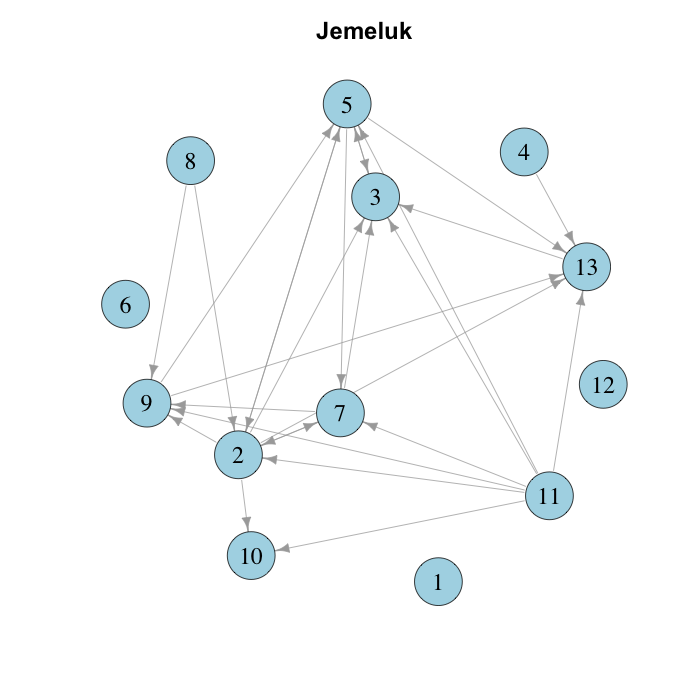}
\end{minipage}
\begin{minipage}[b]{0.3\linewidth}
\centering
\includegraphics[height=1.5in]{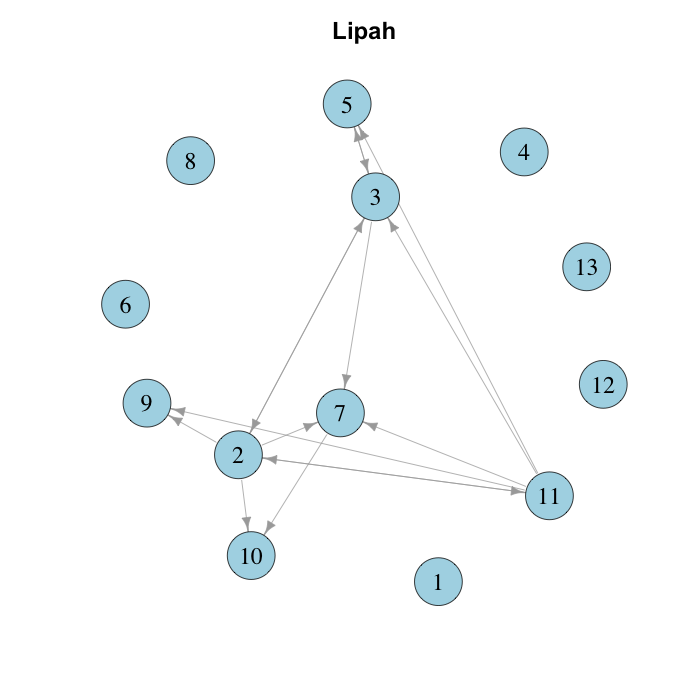}
\end{minipage}
\begin{minipage}[b]{0.3\linewidth}
\centering
\includegraphics[height=1.5in]{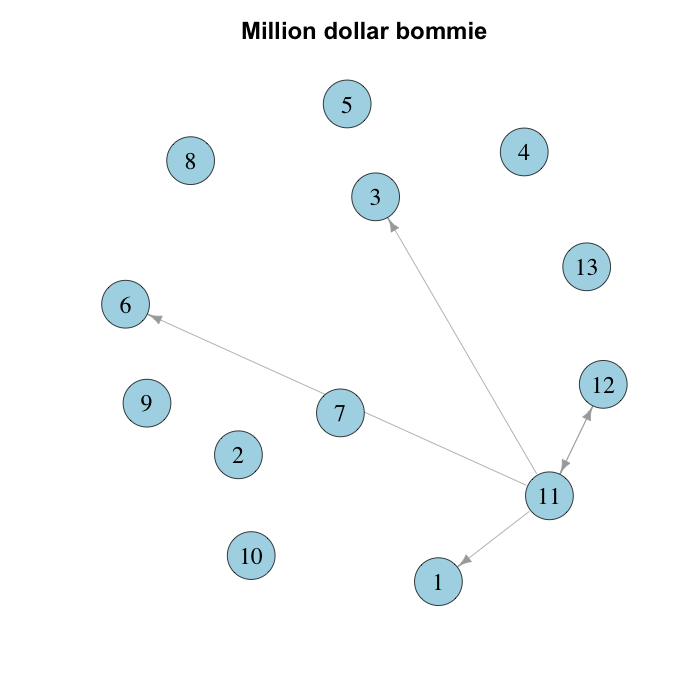}
\end{minipage}
\hfill
\begin{minipage}[b]{0.3\linewidth}
\centering
\includegraphics[height=1.5in]{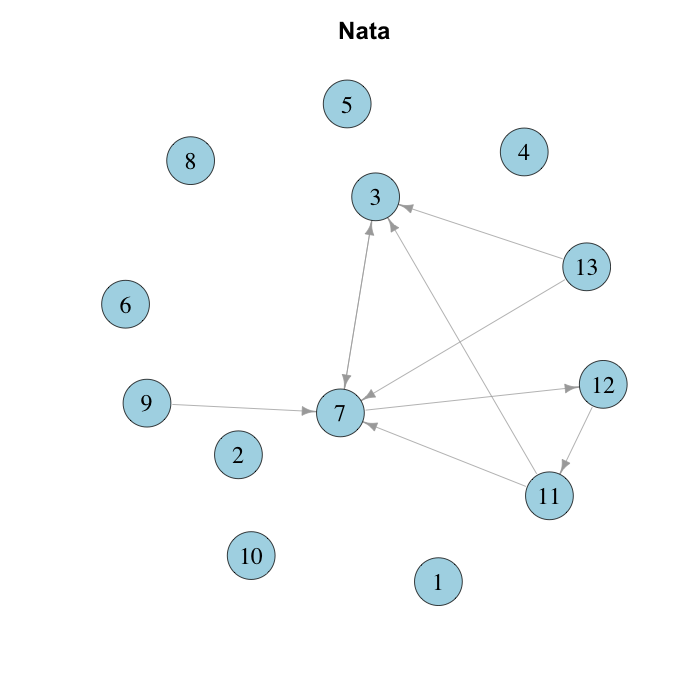}
\end{minipage}
\begin{minipage}[b]{0.3\linewidth}
\centering
\includegraphics[height=1.5in]{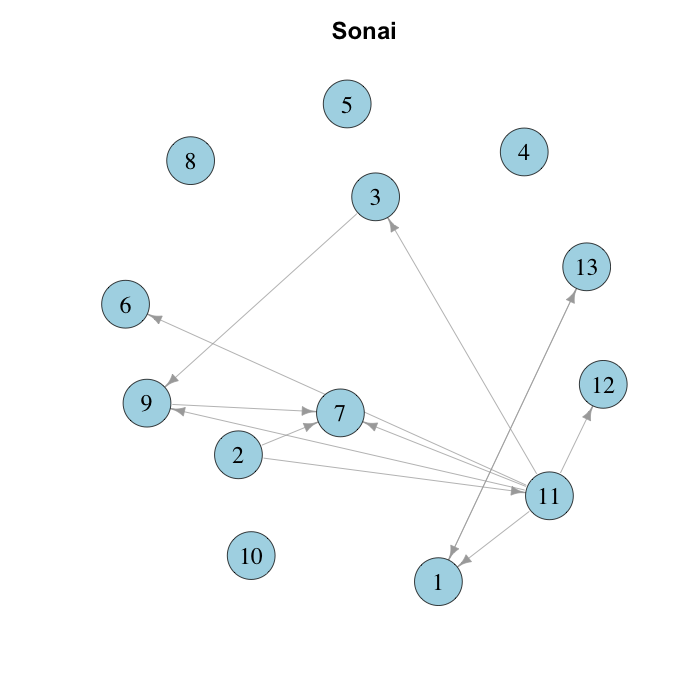}
\end{minipage}
\begin{minipage}[b]{0.3\linewidth}
\centering
\includegraphics[height=1.5in]{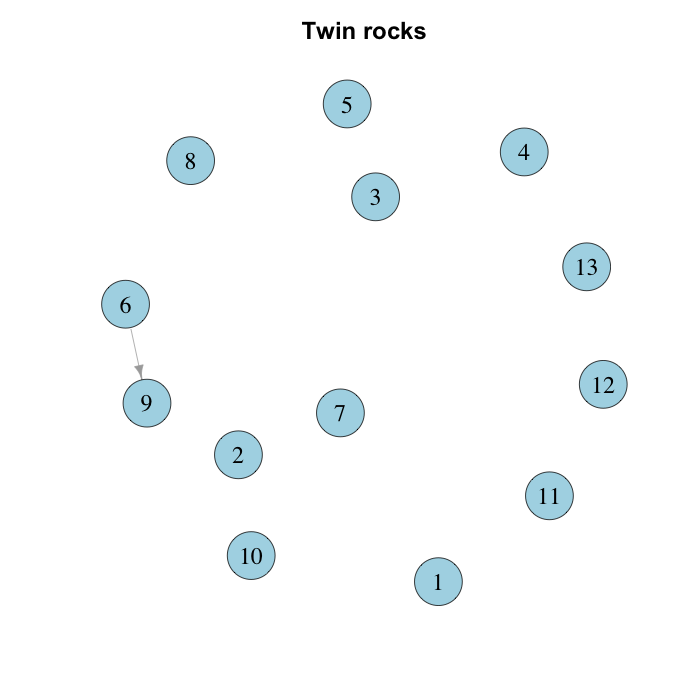}
\end{minipage}
\hfill
\begin{minipage}[b]{\linewidth}
\centering
\includegraphics[height=1.5in]{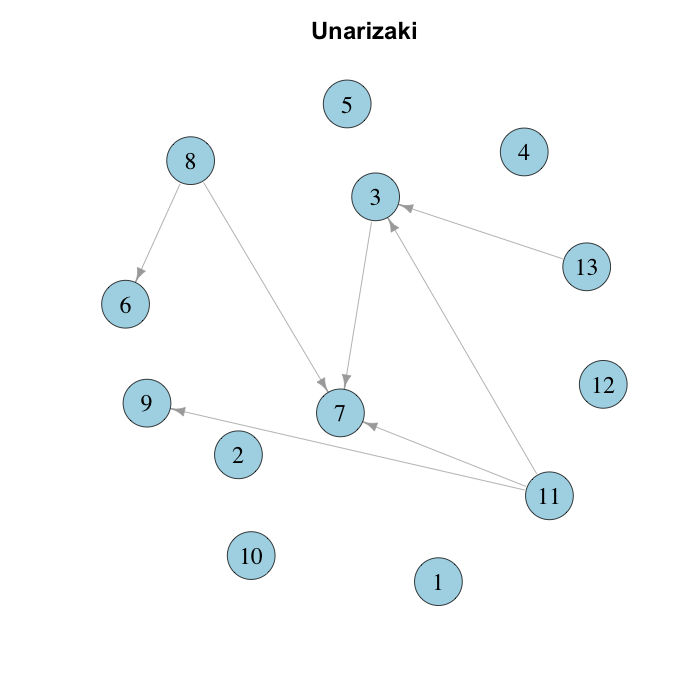}
\end{minipage}
\caption{Network population of fish aggressive interactions with each network representing a reef, at different regions in the Indo-Pacific ocean.}
\label{pop1}
\end{figure}